\documentclass[12pt]{article}
\usepackage{amsmath}
\usepackage{amssymb}
\usepackage{graphicx}
\usepackage{bm}
\usepackage[retainorgcmds]{IEEEtrantools}
\usepackage{bm}
\usepackage{hyperref}
\usepackage[top=3cm, bottom=3.25cm, left=2.75cm, right=2.75cm]{geometry}
\numberwithin{equation}{section}
\newcommand\equ[1] {\begin{equation}#1\end{equation}}
\newcommand\equn[1] {\begin{equation*}#1\end{equation*}}
\newcommand\eqs[1] {\begin{align}#1\end{align}}
\newcommand\eqsn[1] {\begin{align*}#1\end{align*}}
\renewcommand\( {\left(}
\renewcommand\) {\right)}
\newcommand\nn{\nonumber\\}
\newcommand\sign{{\rm sign}}
\newcommand\dy{\rmd\psi}
\newcommand\df{\rmd\phi}
\newcommand\opsi{\omega_\psi}
\usepackage{nicefrac}

\def\beq{\begin{eqnarray}}
\def\eeq{\end{eqnarray}}

\def\bi{\begin{itemize}}
\def\ei{\end{itemize}}

\def\ex#1{{\rm e}^{#1}}
\def\inv{^{\raise.15ex\hbox{${\scriptscriptstyle -}$}\kern-.05em 1}}
\def\tot{_{\raise-.1ex\hbox{${\scriptscriptstyle \text{T}}$}}}
\def\a{\alpha}

\def\o{\omega}

\def\vf{\varphi}
\def\CF{\mathcal F}

\def\del{\partial}
\def\rmd{{\rm d}}

\newcommand\m{\mathcal}
\renewcommand\tilde{\widetilde}

\begin{document}

\begin{titlepage}

\vspace*{2cm}

\begin{center}
{\Huge  \bf Bound states of spinning black holes} \\
\vspace*{0.5cm}
{\Huge  \bf in five dimensions} \\

\vspace*{1cm}

{P. Marcos Crichigno, Flavio Porri, and  Stefan Vandoren}
\bigskip
\bigskip

Institute for Theoretical Physics and Spinoza Institute, Utrecht University\\
Leuvenlaan 4, 3854 CE Utrecht, The Netherlands
\vskip 5mm

\bigskip
\texttt{p.m.crichigno@uu.nl,~f.porri@uu.nl,~S.J.G.Vandoren@uu.nl} \\

\end{center}

\vspace*{1.8cm}

\begin{abstract}

\noindent We find and study supergravity BPS bound states of five-dimensional spinning black holes in asymptotically flat spacetime. These solutions follow from multi-string solutions in six-dimensional minimal supergravity and can be uplifted to F-theory or M-theory. We analyze the regularity conditions and work out the example of a bound state of two black holes in detail. The bound state is supported by fluxes through nontrivial topologies exterior to the horizons and KK momentum. Furthermore, we determine the entropy and compare with other macroscopic BPS solutions. 

\end{abstract}

\end{titlepage}

\section{Introduction}

The phase structure of asymptotically flat five-dimensional BPS objects in supergravity is rich and intricate. For instance, there are spinning black holes \cite{Breckenridge:1996is}, black rings \cite{Elvang:2004rt}, black hole horizons with Lens space topologies \cite{Kunduri:2014kja}, and BPS smooth geometries with no horizon (see  \cite{Gibbons:2013tqa} and references therein). Moreover, there can be bound states between these objects, such as concentric black rings or black saturns \cite{Gauntlett:2004wh,Gauntlett:2004qy}, or a bound state of a black hole with smooth centers outside the black hole horizon. For a review, see e.g. \cite{Bena:2007kg}. In this paper, we investigate the possibility of having bound states of spinning black holes, where each of the black holes has an $S^3$ horizon topology. 

All these objects can be realized microscopically in string theory. The case with least supersymmetry is M-theory on a Calabi-Yau threefold $X$, or F-theory on $X\times S^1$ for elliptic $X$. When an F-theory picture is available, a 5d black hole arises from a 6d black string by wrapping a D3-brane over ${\cal C} \times S^1$, where $\cal C$ is a curve in the base of $X$ \cite{Vafa:1997gr}. Multicenter bound states of spinning black holes arise when ${\cal C}$ degenerates into multiple curve components of lower genus \cite{Haghighat:2015ega}. This work served as a motivation for the present study. Since the F-theory picture describes the black holes as 6d black strings wrapped over a circle, we are also led to investigate the question of the existence of multicentered black strings in six dimensions. These multicentered string configurations can also be constructed from type IIB compactifications on $K3$ or $T^4$. Such centers might bind or not in spacetime, and we derive the conditions in 6d supergravity for them to form regular BPS bound states. Upon reducing on $S^1$, they describe black hole bound states in five dimensions. Our analysis will be done in minimal (1,0) supergravity, which has an F-theory lift in terms of elliptically fibered $X$ with base $\mathbb{P}^2$. In 5d, the bound state black hole system will therefore be a BPS solution of 5d supergravity coupled to one vector multiplet.\footnote{\label{foot1}Of course, there are also BPS black hole solutions in minimal 5d supergravity \cite{Gauntlett:1998fz,Gauntlett:2002nw}, but these do not uplift to F-theory. Therefore, there is no underlying microscopic description in terms of a 6d black string with an AdS$_3$ near horizon factor whose entropy is governed by a CFT$_2$.} We expect these bound states to persist in the presence of additional matter multiplets. This would be relevant for string compactifications with more supersymmetry, such as M-theory on $T^6$ or IIB on $T^5$.

One of the questions we address is whether regular, multicenter solutions exist in regions of parameter space where the single-center solutions would violate the cosmic censorship bound (CCB). This was addressed in the case in which only one center has a finite-size horizon in \cite{Bena:2011zw,Kunduri:2014kja,Kunduri:2014iga}. However, to the best of our knowledge, a detailed analysis of the case with multiple horizons has not been carried out.  Although the local form of the solutions is known, there are various regularity conditions that must be imposed on the metric for the solution to be a good background, leading to a set of nontrivial constraints on the parameters describing the local solution. Whether there is a nonzero space of solutions to these constraints requires a careful analysis. 

In this paper, we carry out this analysis in the case of a solution describing two identical finite-size spinning black holes, and a smooth center.  The space contains two topological two-cycles connecting the black holes to the smooth center, and the whole system is bound by a nonzero flux through these cycles. We find that there is a narrow region around the CCB where these configurations exist and are everywhere regular. Moreover, their entropy dominates over the single-centered black hole in a small subregion where both exist.

We begin our analysis with a description of multi-string solutions in six dimensions in Section 2. In Section 3 we reduce over a circle to five dimensions and in Section 4 we discuss the bound states of 5d black holes. We end with some discussion in Section 5, and give some technical details in the Appendices. 

\section{Multiple strings in 6d}
\label{Bound strings in 6d}

Consider F-theory on an elliptically fibered Calabi-Yau threefold $X$ with base $B$ \cite{Vafa:1996xn,Morrison:1996na,Morrison:1996pp}. In six dimensions this gives rise to the Poincar\'e multiplet (containg the graviton and a self-dual tensor), $n_T=h^{1,1}(B)-1$ tensor multiplets (with anti-selfdual tensors), and $n_V=h^{1,1}(X)-h^{1,1}(B)-1$ vector multiplets. The tensors descend from the RR-four form 
\begin{equation}
C_{(4)}=\sum_{i=1}^{h^{1,1}(B)} C_{(2)}^i \wedge \alpha_i\ ,\qquad C_{(2)}^i=\int_{\gamma^i} C_{(4)}\ ,
\end{equation}
where $\alpha_i$ and $\gamma^i$ form a basis of harmonic (1,1)-forms and dual two-cycles respectively. 
There are also $n_H=h^{2,1}(X)$ hypermulitplets from the complex structure deformations of $X$ but they play no role in our analysis and are frozen to constant values. The simplest setup for the study of black holes is when the base is chosen to be $B=\mathbb{P}^2$, for which there are no vector multiplets and no tensor multiplets, so we are led to pure minimal chiral (1,0) supergravity in six dimensions, with bosonic fields the metric and a self-dual three-form $\widehat{G}={\rm d}C_{(2)}$. The BPS equations in this supergravity theory were studied and analyzed in \cite{Gutowski:2003rg}. 

BPS black strings in six dimensions arise from wrapping D3-branes in F-theory over a curve ${\cal C}$ in the base $B$ \cite{Vafa:1997gr,Haghighat:2015ega,Bena:2006qm}. The near-horizon geometry of such a string in six dimensions is AdS$_3\times S^3$. For $B=\mathbb{P}^2$, we have only one K\"ahler class $[H]$ and thus $[{\cal C}]=d[H]$ for integer $d$, the degree of the curve. Wrapping the string over an $S^1$ with $N$ units of momentum yields a 5d black hole with entropy \cite{Breckenridge:1996is}
\begin{equation}\label{F theory entropy}
S=2\pi{\sqrt {\frac{d^2 N}{2}-J^2}}\ ,
\end{equation}
where $J$ is the angular momentum of the black hole.

Moving within the class, the curve ${\cal C}$ can degenerate into $n_C$ curve components and multi-string branches with $d=d_1+d_2+...+d_{n_C}$ can arise \cite{Haghighat:2015ega}. These multiple strings may or may not bind in spacetime; we discuss the conditions under which they form BPS bound states in subsection~\ref{Bound states}. 
\subsection{Supersymmetric solutions in 6d}

The bosonic content of six-dimensional minimal supergravity is the graviton $\hat g_{\hat \mu\hat \nu}$ and a self-dual three-form field $\widehat G_{\hat \mu\hat \nu \hat \rho}$. All supersymmetric solutions of minimal supergravity in 6d were described in \cite{Gutowski:2003rg} (for subsequent work see, e.g., \cite{Bena:2011dd}), which we closely follow below. The metric is given by\footnote{Here we adopt the signature convention mostly plus, unlike \cite{Gutowski:2003rg}, and redefined the function $\mathcal F_{here}=-\mathcal F_{there}$. }
\equ{\label{6d metric general} 
{\rm d}s^{2}_{6}= -2 H^{-1}({\rm d}u+\beta)\({\rm d}v+\omega-\frac{\mathcal F}{2}({\rm d}u+\beta)\)+H\, {\rm d}s^{2}_{\text{HK}_{4}}\,,
}
where ${\rm d}s^{2}_{\text{HK}_{4}}$ is a four-dimensional  hyperk\"ahler base and $\beta, \omega$ are 1-forms on $ \text{HK}_{4}$. The vector field $\partial_{v}$ is a null Killing vector field. In full generality $\mathcal F=\mathcal F(u,x)$, $H=H(u,x)$, where $x$ are coordinates in the base, but
here we consider the case in which $\partial_{u}$ is a (spacelike) Killing vector, i.e.,
\equ{
\mathcal F=\mathcal F(x)\,,\qquad  H=H(x)\,.
}
In this case the 3-form field is given by
\equ{\label{3-form in 6d}
\widehat G=\frac12 \ast_{4}{\rm d}H-\frac12 e^{+}\wedge ({\rm d}\omega)^{-}+\frac12 H^{-1}e^{-}\wedge {\rm d}\beta-\frac12 e^{+}\wedge e^{-}\wedge H^{-1}{\rm d}H\,,
}
where $$e^{+}=H^{-1}({\rm d}u +\beta)\,,\quad e^{-}={\rm d}v+\omega -\tfrac{\m F}{2}({\rm d}u+\beta)\,,\quad  {\rm d}\omega^{-}=\tfrac12({\rm d}\omega-\ast_{4}\rm d\omega)~.$$

For these $u$-independent solutions, one may take the $u$-coordinate to be periodic, $u\sim u+\ell$, thus the total F-theory geometry is
\begin{equation}
{\mathbb{R}}\times {\text{HK}_{4}}\times S^1\times CY_3\ .
\end{equation}

Within this family of solutions, an interesting class is when HK$_{4}$ is taken to be a multicenter Gibbons-Hawking (GH) space \cite{Gibbons:1979zt}, whose metric has the form of a $U(1)$ fibration over flat $\mathbb R^{3}$:
\equ{\label{gibbons hawking metric}
\rmd s_{\text{HK}_{4}}^2 = H_2\inv \(\dy + \bm \chi\)^{2} + H_2 \, \rmd s^2_{\mathbb R^3}~,
} 
with $H_{2}$ the harmonic function on $\mathbb{R}^{3}$: 
\equ{
\label{H2 and eq chi}
H_{2}=m_{\infty}+\sum_{a} \frac{m_{a}}{|\vec x -\vec x_{a}|}\,,\, \qquad \ast_3 \rmd\bm \chi = \rmd H_2\,,
}
where $m_{a}$, $a=1,...,N$ are integers. The coordinates in $\mathbb R^3$ are $(r,\theta,\phi)$, where $\theta\in[0,\pi]$ and $\phi\in[0,2\pi]$ are coordinates on the round $S^{2}$ and $\psi\in[0,4\pi]$ is the fiber direction. \\ 

Before we proceed describing the full solution, let us review some well known facts about the Gibbons-Hawking metrics  \eqref{gibbons hawking metric} that will be useful later. Close to a center $\vec x \to \vec x_{a}$, the metric becomes $\mathbb{R}^{4}/\mathbb{Z}_{|m_{a}|}$. In particular, this means that $m_{a}\in \mathbb{Z}$ and for $|m_{a}|=1$ the metric is locally $\mathbb{R}^{4}$. Another important property is that since the $\psi$-fiber shrinks to zero size as one approaches any of the centers with $m_{a}\neq 0$, there is a nontrivial topological 2-cycle between any two centers, spanned by $\psi$ and any curve connecting them. Although there are $\frac12 N(N-1)$ number of such cycles, there are a total of $N-1$ independent 2-cycles in homology. The asymptotics of these spaces depend on the value of $m_{\infty}$. For $m_{\infty}=0$ the space is known as multi Eguchi-Hanson, with  asymptotics $\mathbb{R}^{4}/\mathbb{Z}_{|m\tot|}$, where $m\tot=\sum_{a}m_{a}$. For a single center with $m=1$ the metric is simply flat  $\mathbb R^{4}$. 

The multicenter case with $m_{\infty}\neq 0$ is known as multi Taub-NUT, with asymptotics $\mathbb{R}^{3}\times S^{1}$. In this case one may reduce further along this asymptotic $S^{1}$ to four dimensions. We will comment briefly on this in subsection 3.5.

It should be noted that for standard Gibbons-Hawking metrics  one requires $H_{2}>0$ (and hence $m_{a}>0$) to ensure the metric is positive definite. However, this condition can be relaxed here as long as the warp factor $H$ in \eqref{6d metric general} compensates for the sign change in $H_{2}$ and the full 6d metric has the correct signature. Indeed, we will be particularly interested in configurations where some $m_{a}=-1$. These metrics were used to construct microstates geometries in 5d \cite{Bena:2005va,Berglund:2005vb} and 6d \cite{Saxena:2005uk} and are often referred to as {\it ambipolar} Gibbons-Hawking metrics.   \\

We now return to the description of the full solution. Assuming that the Killing vector  $\partial_{\psi}$ of the GH base  extends to a symmetry of the full space, the complete supergravity background is then given in terms of five additional generic harmonic functions on $\mathbb{R}^{3}$ (see \cite{Gutowski:2003rg} for more details)
 \eqs{\label{Hs}
H_{1}&=\mu_{\infty}+\sum_{a} \frac{\mu_{a}}{|\vec x -\vec x_{a}|}\,, \qquad H_{3}=q_{\infty}+\sum_{a} \frac{q_{a}}{|\vec x -\vec x_{a}|}\,,\\ \nonumber
H_{4}&=p_{\infty}+\sum_{a} \frac{p_{a}}{|\vec x -\vec x_{a}|}\,,  \qquad H_{5}=n_{\infty}+\sum_{a} \frac{n_{a}}{|\vec x -\vec x_{a}|}\,, \qquad H_{6}=j_{\infty}+\sum_{a} \frac{j_{a}}{|\vec x -\vec x_{a}|}\,.
}
\begin{figure}[tbp]
\centering
\includegraphics[page=2]{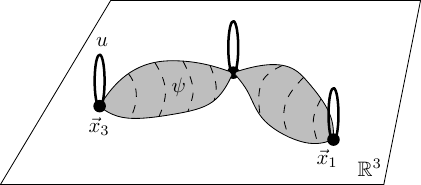}
\caption{A multi-string configuration in 6d. The distances between the strings are constrained by the integrability equations \eqref{bubble eqs}. The base here is a 3-center Gibbons-Hawking space.}
\label{strings}
\end{figure}
In what follows we adopt a notation in which $r \equiv |\vec x|$, $r_a \equiv |\vec x - \vec x_a|$,  $r_{ab} \equiv |\vec x_{a} - \vec x_b|$ and denote the set of harmonic functions by
\equ{
\mathbb H=(H_{1},H_{2},H_{3},H_{4},H_{5},H_{6})\,.
 }
The functions $\mathcal F, H$ appearing in the metric are given in terms of these by
\equ{\label{F and H general}
\mathcal F=H_{5}+H_{2}^{-1}H_{4}^{2}\,, \qquad H= H_{1}+H_{2}^{-1}H_{3}H_{4}\,,
}
and the 1-forms  $\beta= \beta_\psi \(\dy + \chi\) + \bm\beta_{i} \,\rmd x^i$ and  $\omega= \omega_\psi \(\dy + \chi\) + \bm\omega_i \,\rmd x^i$  are given by
\begin{align}\label{eqs beta}
\beta_\psi &= H_2\inv H_3\,, && \ast_3 \rmd\bm\beta = -\rmd H_3~,\\ \label{eqs omega}
 \opsi &= H_2^{-2} H_3 H_4^2 + H_2\inv\(H_1 H_4 + \tfrac12 H_5 H_3 \) + H_6\,, && 
 \ast_3 \rmd\bm\omega =  \langle \mathbb H,\rmd \mathbb H\rangle ~,
\end{align}
where  we introduced the symplectic product $\langle u,v\rangle\equiv u^{\top}\,\Omega \, v$, with
\equ{\label{matrix omega}
\Omega=\left(
\begin{array}{cccccc}
 0 & 0 & 0 & -1 & 0 & 0 \\
 0 & 0 & 0 & 0 & 0 & 1 \\
 0 & 0 & 0 & 0 & \frac{1}{2} & 0 \\
 1 & 0 & 0 & 0 & 0 & 0 \\
 0 & 0 & -\frac{1}{2} & 0 & 0 & 0 \\
 0 & -1 & 0 & 0 & 0 & 0 \\
\end{array}
\right)\,.
}
It is convenient to denote the residues and constant parts in $\mathbb H$ by the vectors
\equ{
 \Gamma_{a}=(\mu_{a},m_{a},q_{a},p_{a},n_{a},j_{a})\,, \qquad \Gamma_{\infty}=(\mu_{\infty},m_{\infty},q_{\infty},p_{\infty},n_{\infty},j_{\infty})\,,
}
The symplectic pairing acts naturally on these  by
\equ{\label{prod gammas}
\langle \Gamma_{a}, \Gamma_{b} \rangle= p_{a}\mu_{b}-\mu_{a}p_{b}+m_{a}j_{b}-m_{b}j_{a}+\frac12\(q_{a}n_{b}-q_{b}n_{a}\)\,.
}

As shown in \cite{Gutowski:2003rg}, the solution reviewed above is the most general $u$-independent solution with a Gibbons-Hawking base space whose Killing vector field $\partial_{\psi}$ extends to a symmetry of the full solution. Specifying a particular local solution in this class amounts to giving a set of locations $\{\vec x_{a}\}$ of the poles in the harmonic functions, their residues $\{\Gamma_{a}\}$ and the asymptotic values $\Gamma_{\infty}$.  To write the solution explicitly one must determine the 1-forms $\bm \chi, \bm \beta,\bm \o$ from equations \eqref{H2 and eq chi},  \eqref{eqs beta} and \eqref{eqs omega}, respectively. The expression for these is given in  Appendix~\ref{axysimmetric solution} in a simplified case where all the GH centers lie on a single line inside $\Bbb R^{3}$.

The asymptotics of the metric \eqref{6d metric general} is controlled by the asymptotics of the GH base and the behavior of the functions $H,\mathcal F$ and 1-forms $\o, \beta$ as $r\to \infty$. Setting $m_{\infty}=0$ and requiring that asymptotically $H, \mathcal F\to 1$ and $\o, \beta \to0$, the metric asymptotes to $\Bbb R^{1,1}\times \Bbb R^{4}/\Bbb Z_{m\tot}$, where $m\tot\equiv \sum_{a}m_{a}$ must be positive for the metric to have the correct signature. See Appendix~\ref{Asymptotics 6d app} for details.

\paragraph{Black strings and black tubes.}

The single black string solution is obtained by taking the harmonic functions \eqref{H2 and eq chi} and \eqref{Hs} to have a single pole at the origin with residues $\Gamma=(\mu,m,q,p,n,j)$, with $m$ positive integer\footnote{For $m\neq1$ the spacetime is a $\mathbb Z_m$ orbifold  which can be undone choosing $\psi$ to have period $4\pi m$. We do not do this here since we want to obtain an entropy formula for a generic set of charges.} and constant parts $\Gamma_{\infty}$ given by \eqref{gamma inf r4 app}. The metric \eqref{6d metric general} then reads
\begin{align}\label{black string metric}
\rmd s^2= -2\left(1+\frac{\tilde Q}{4\sqrt2m\,r}\right)^{-1} \left[\rmd v+\frac{J_\psi}{8m^2r}\left(\dy+m\cos\theta\,\df\right) - \frac12 \left(1+\frac{Q}{4m\,r}\right) \rmd u' \right] \rmd u'  \nn
+ \left(1+\frac{\tilde Q}{4\sqrt2m\,r}\right)\left[\frac r m \left(\dy+m\cos\theta\,\df\right) ^2 +m\frac{\rmd r^2}{r} + m\,r\,\rmd\Omega_2^2   \right]\,,
\end{align}
where we defined $u' = u + \frac q m \psi$. Since $u$ and $\psi$ are both periodic, we must impose the quantization condition $\frac{4\pi}\ell\frac{q}{m}\in\mathbb Z$ in order for the reparametrization to be globally defined. We also introduced the combinations
\equ{\label{charges single string}
\tilde{Q} =4\sqrt2(\mu m +q p) \,, \qquad  Q  = 4(n m +p ^{2})\,,\qquad J_{\psi}=8(q p^{2} +\mu  p m +\frac{q n }{2}m +j m ^{2})\,.
}
The spacetime \eqref{black string metric} is asymptotically $S^1\times {\mathbb R}^{1,4}/\mathbb Z_{m}$ where the $S^1$ is parametrized by $u'$ and has an event horizon at $r=0$. The near-horizon geometry\footnote{A solution which is globally BTZ$\times S^{3}$ can be obtained by taking a single center and setting $\Gamma_{\infty}=0$. For other six-dimensional solutions that asymptote to AdS$_{3}\times S^{3}$ see e.g. \cite{Bobev:2012af}.} is a direct product of an extremal BTZ black hole and a round $S^3/\mathbb Z_{m}$. The Bekenstein-Hawking entropy associated to the black string horizon reads
\equ{\label{black string area}
S = \frac{\text{Area}}{4\,G_6} =  2 \pi \sqrt{\frac{Q\tilde Q^2}{2m^{2}} - \frac{J_\psi^2}{m^{2}}}\,,
}
where we chose conventions  \cite{Bena:2006qm}  in which $G_6 = \frac{\ell\pi}{4}$, with $\ell$ the period of $u'$. Setting $m=1$ and with the identification  with F-theory quantities $J=J_{\psi},\, N=Q,\,  d=\tilde Q$, this matches the entropy  \eqref{F theory entropy}. Since the F-theory quantities are quantized in the microscopic theory, the charges $J_{\psi},Q,\tilde Q$ are integers, which explains our choice of normalizations in \eqref{charges single string}.

The extension to the multi-string case is straightforward. Taking a generic configurations with charge vectors $\{\Gamma_a\}$  the metric close to any center with  $m_{a}\neq 0$ will resemble the $r\to0$ limit of \eqref{black string metric} with coefficients $\tilde Q_a$, $Q_a$ and $J_\psi{}_a$. In fact, unlike the case of a single string, one may (and we will) allow for some $m_{a}<0$ as long as the asymptotic condition $m\tot>0$ is satisfied. The entropy of each string is given by \eqref{black string area} with the corresponding charges.

Although not the focus of the present paper, another interesting possibility is when $m_a=0$ for some centers.  In this case the topology of the horizon at $\vec x=\vec x_a$ degenerates into
\equ{
S^1_u\times S^3/\mathbb Z_{|m_{a}|} \to S^1_u\times S^1\times S^2~,
}
and the object is then a (circular) {\it black tube} rather than a black string. This is nothing but the 6d uplift of the supersymmetric black ring \cite{Elvang:2004rt}. Although the entropy formula \eqref{black string area} might look singular, it is still valid in the $m_{a}=0$ case by taking the limit (see expressions \ref{charges single string}). 

We will assume $m_a\neq0$ for the rest of the paper since we are interested in black strings (black holes) in 6d (5d). Next, we discuss the conditions for these objects to form bound states.

\subsection{Bubble equations}
\label{Bound states}

Although any set of locations $\vec x_{a}$ of the strings in $\mathbb{R}^{3}$ provide local supergravity solutions,  these will typically have Dirac string-like  singularities. In similar settings in four  \cite{Denef:2000nb} and five dimensions \cite{Bena:2005va}, it is well known that imposing the absence of such singularities leads to a constraint on the relative locations of the GH centers. This is also the case in six dimensions (see e.g. \cite{Bena:2008wt}). As in the lower-dimensional cases, this arises from requiring the 1-form $\bm \omega$ appearing in the metric to be globally defined, which implies 
\equ{
\rm d^{2}\bm \omega =0\,.
}
Taking $\rmd\ast_{3}$ on both sides of \eqref{eqs omega}  leads to
\equ{\label{bubble eqs}
\sum_{b\neq a}\frac{\langle \Gamma_{a}, \Gamma_{b} \rangle}{r_{ab}}=\langle \Gamma_{\infty},\Gamma_{a}\rangle\,, \qquad \qquad a=1,...,N\,.
}
These equations impose constraints on the relative distances $r_{ab}$ in $\mathbb{R}^{3}$ of the GH centers and their charges. These are usually referred to as ``bubble equations,'' because they control the size of the ``bubbles," or 2-cycles, in the Gibbons-Hawking base (see e.g.  \cite{Bena:2007kg}).  In a similar setting in four dimensions \cite{Denef:2000nb}, they are referred to as ``integrability equations.''

We note that summing over $a$ on both sides of (\ref{bubble eqs}), the left-hand side vanishes identically and thus a consistency requirement  is 
\equ{\label{sumbubb}
\sum_{a}\langle \Gamma_{\infty},\Gamma_{a}\rangle\equiv\langle \Gamma_{\infty},\Gamma\tot\rangle=0\,,
}
which can be interpreted as the condition that there are no Dirac strings running to infinity.

The constraint \eqref{bubble eqs} coincides with the bubble equations found in five dimensions \cite{Bena:2005va}. This is not hard to explain. As we will discuss in section \ref{reduction to five}, a 5d BPS solution (in the time-like class) is obtained from the 6d solution by reducing along the direction  $u\sim u+\ell$.  Since the bubble equations depend only on the GH base, which is unaffected by the dimensional reduction, these coincide in 5d and 6d. 
\subsection{Dualities and Charges}
\label{Dualities and Charges}

As reviewed above, the class of solutions considered in this paper is characterized by the six harmonic functions $\mathbb H=(H_{1},H_{2},H_{3},H_{4},H_{5},H_{6})$. 
Since these are generic harmonic functions, and a linear combination of harmonic functions is harmonic, it is clear that sending $\mathbb H\to g \, \mathbb H$ with $g\in GL(6,\mathbb{R})$ will send a solution to a solution.  An interesting question is whether this operation preserves regularity, in particular the absence of Dirac-string singularities. One way to ensure this is if the bubble equations \eqref{bubble eqs} are preserved, which leads us to consider the subgroup, $Sp(6,\mathbb{R})$,  preserving the symplectic product, i.e., $g^{\top}\Omega g=\Omega$.

One example of such transformations is given by the two-parameter set of transformations
\equ{\label{gauge symmetry}
g_{gauge}=\begin{pmatrix}
1& -g_{1}g_{2} &-g_{1} &-g_{2} &0 &0 \\
0& 1& 0&0 &0 &0 \\
0& g_{2}& 1& 0&0 &0 \\
0& g_{1}&0 &1 &0&0 \\
0& -g_{1}^{2}&0 & -2g_{1}&1 & 0\\
-g_{1}&\frac12 g_{2}g_{1}^{2} &\frac12 g_{1}^{2} &g_{1}g_{2} &-\frac12g_{2} &1 \\
\end{pmatrix}\,,
}
where $g_{1,2}$ are real parameters. In fact, these transformations form a two-dimensional subgroup of $Sp(6,\mathbb{R})$ and it is easy to see that they 
leave the functions $H, \mathcal F$, as well as the 1-form $\omega$ invariant. The 1-form $\beta$ transforms by an exact term: $\beta \to \beta-g_{2}\,\rmd\psi$, which can be undone by the coordinate transformation $u\to u+g_{2}\psi$. Furthermore, since the function $H_2$ is invariant, the effect of the transformation \eqref{gauge symmetry} is a simple, unphysical, change of coordinates. The explicit action on the residues reads
\eqs{\nonumber
&\mu_{a}\to \mu_{a}-g_{1}q_{a}-g_{2}p_{a}-g_{1}g_{2}m_{a}\,,\qquad m_{a}\to m_{a}\,, \qquad q_{a}\to q_{a}+g_{2}m_{a}\,,\\ \label{shift residues}
&p_{a}\to p_{a}+g_{1}m_{a}\,,\qquad n_{a}\to n_{a}-2g_{1}p_{a}-g_{1}^{2}m_{a}\,,\\ \nonumber
& j_{a}\to j_{a}-g_{1}\mu_{a}-\frac12 g_{2}n_{a}+g_{1}g_{2}p_{a}+\frac12 g_{1}^{2}q_{a}+\frac12g_{2}g_{1}^{2}m_{a}\,.
}
In particular, we note that one may always set the $p_{a},q_{a}$ charges of one center to zero by choosing $g_{1},g_{2}$ appropriately (provided  $m_{a}\neq 0$). In the context of M-theory on $T^{6}$ these are referred to as ``gauge'' transformations \cite{Bena:2005ni}. This symmetry can  be used to construct physically relevant combinations of the residues, as we discuss next.

For an  $N$ number of centers, there are a total of $6N$ residues in the harmonic functions.\footnote{In principle there are six more parameters in $\Gamma_\infty$ which should be considered in the counting. In fact, one can construct gauge-invariant combinations analogous to \eqref{gauge inv quant}. However, the latter vanish in the asymptotically flat solutions considered here.} Due to  the redundancy we just described, the residues themselves are not physical quantities. Since the redundancy is characterized by two parameters $g_{1,2}$ there should be a total of $6N-2$ gauge-invariant combinations. These are the $m_{a}$'s themselves, together with the $5N-2$ independent combinations
\eqs{\nonumber
&\tilde{Q}_{a}\equiv 4\sqrt2(\mu_{a}m_{a}+q_{a}p_{a})\,, \qquad  Q_{a} \equiv 4(n_{a}m_{a}+p_{a}^{2})\,,\\ \label{gauge inv quant}
&J_{a}^{\psi}\equiv 8(q_{a}p^{2}_{a}+\mu_{a} p_{a}m_{a}+\frac{q_{a}n_{a}}{2}m_{a}+j_{a}m_{a}^{2})\,,\\ \nonumber
&f_{a,a+1}\equiv \frac{q_{a+1}}{m_{a+1}} - \frac{q_{a}}{m_{a}}\,, \qquad  \widetilde{f}_{a,a+1}\equiv \sqrt2\(\frac{p_{a+1}}{m_{a+1}} - \frac{p_{a}}{m_{a}}\)\,.
}
In the last line we have assumed $m_a\neq0$ for all $a$. The quantities above have a clear physical interpretation, as we will discuss in section~\ref{reduction to five}. In the case of a single GH center, these quantities reduce exactly to the quantities \eqref{charges single string}.

Another interesting subgroup of $Sp(6,\mathbb{R})$ is given by
\equ{\label{definition Msf}
g_{\text{SF}}=\begin{pmatrix}
1& 0& 0&0 &0 &-2\gamma_{1} \\
-2\gamma_{1}\gamma_{2}& 1& \gamma_{2}&2 \gamma_{1} &-\gamma_{1}^{2} &2\gamma_{1}^{2}\gamma_{2}\\
-2\gamma_{1}& 0& 1& 0&0 &2\gamma_{1}^{2} \\
-\gamma_{2}& 0&0 &1 &-\gamma_{1}&2\gamma_{1}\gamma_{2} \\
0& 0&0 & 0&1 & -2\gamma_{2}\\
0&0 &0 &0&0&1 \\
\end{pmatrix}\,,
}
where $\gamma_{1,2}$ are real parameters. These transformations do not leave the quantities \eqref{gauge inv quant} invariant. Instead, they lead to a new solution, characterized by the transformed quantities $\{m_{a}',{Q}_{a}', \tilde Q_{a}', J_{a}^{\psi\, '},f_{a,a+1}', \widetilde{f}_{a,a+1}'\}$. Although the transformation acts non-trivially on these quantities one can see that the entropy \eqref{black string area} is invariant. Since the transformation preserves the bubble equations, the new supergravity background will necessarily be free of Dirac singularities, provided the original background is. In the case of M-theory on $T^{6}$, these correspond to a subgroup of the $E_{7(7)}$ U-duality group, referred to as generalized spectral flow transformations in \cite{Bena:2008wt}.

Finally, another subgroup of $Sp(6,\mathbb{R})$ leaving the entropy invariant is given by 
\equ{\label{sp rescaling}
g_{resc.}=\left(
\begin{array}{cccccc}
\beta_{1} & 0 & 0 & 0 & 0 & 0 \\
 0 & \beta_{2} & 0 & 0 & 0 & 0 \\
 0 & 0 & \beta_{1}^{2}\beta_{2}& 0 & 0 & 0 \\
 0 & 0 & 0 & \beta_{1}^{-1} & 0 & 0 \\
 0 & 0 & 0 & 0 & \beta_{1}^{-2}\beta_{2}^{-1} & 0 \\
 0 & 0 & 0 & 0 & 0 & \beta_{2}^{-1} 
\end{array}
\right)\,,
}
with $\beta_{1,2}$ real parameters, which acts by a simple rescaling of the residues. \\

Before we proceed we make a brief comment. We have shown that the particular $Sp(6,\mathbb{R})$ group elements (\ref{gauge symmetry}),(\ref{definition Msf}) and \eqref{sp rescaling} leave the entropy invariant. In fact it is not difficult to show that combinations of these transformations form the most general subgroup $H\subset Sp(6,\mathbb{R})$ with this property and that they form the direct product  
\equn{
H= SL(2,\mathbb{R})\times SL(2,\mathbb{R})\,,
}
where each factor is a combination of all three types of transformations described above. The meaning of the $Sp(6,\mathbb{R})$ group and its consequences will be studied in \cite{CPV}.

\section{Reduction to five dimensions}
\label{reduction to five}

Compactification of F-theory on $X\times S^1$ yields an effectively five-dimensional theory which is the circle reduction of the 6d theory we considered in the previous section. By wrapping the black strings over $S^1_u$ with quantized momenta, we obtain charged spinning black holes in five dimensions. In this section, we discuss the dimensional reduction from 6d to 5d. 

Dimensional reduction of 6d minimal supergravity to 5d leads to the following bosonic fields: a metric $g_{\mu\nu}$, a scalar $\varphi$, and two vector fields $A_{\mu}$ and $\tilde A_{\mu}$. For further details we refer the reader to Appendix \ref{reduction}. Reduction of the 6d metric and three-form along the $u$-direction\footnote{Another possibility would be to reduce along the GH fiber $\dy$, in this way one can obtain 5d solutions of the ``null" class \cite{Gauntlett:2002nw}. We do not discuss this here.} yields
\eqs{\nonumber 
\rmd s^2_6 &= \ex{2\vf} \(\rmd u + A\)^2 + \ex{-2\vf/3}\, \rmd s^2_5\,,\\ \label{kk6-5}
\widehat G&=G+\frac12\rmd \widetilde A\wedge \(\rmd u + A\)\,,
}
with the five-dimensional, Einstein-frame metric
\equ{\label{BPSmetric5d}
\rmd s^2_{5}= -f^2 \(\rmd t +\omega\)^2 + f\inv \rmd s_{\text{HK}_{4}}^2\,, \qquad \qquad f\inv =\(H^{2} \mathcal F\)^{1/3}\,,
}
where we relabeled $\rmd t =\rmd v$. For BPS solutions of the Gibbons-Hawking type, the radius and the two vector fields read
\begin{align} \nonumber
e^{2\varphi} &= H^{-1}\mathcal F\,,		\\ 
A &= - \(H_5 + H_2\inv H_4^2\)\inv \( \rmd t +\omega\) + \beta\,,\label{bos fields 5d app}	\\
 \tilde A &= - \(H_1 + H_2\inv H_3 H_4\)\inv \( \rmd t +\omega\) + \gamma	\,, \nonumber
\end{align}
where $\beta$ satisfies \eqref{eqs beta} and  $\gamma= \gamma_\psi \(\dy + \chi\) + \bm\gamma_{i} \,\rmd x^i$ similarly satisfies
\begin{align}
\gamma_\psi = H_2\inv H_4\,, \qquad \ast_3 \rmd\bm\gamma = -\rmd H_4\,.
\end{align}
These solutions---and their extensions with an arbitrary number of vector multiplets---were studied in  \cite{Gauntlett:2004qy}.

Note that the five-dimensional solution is still described by six harmonic functions and the residues and locations of the centers are still constrained by the same bubble equations \eqref{bubble eqs}. One then concludes that the five-dimensional solution describes a bound state of black holes if and only if the six-dimensional solution corresponds to a bound state of black strings. Before discussing the variety of possible five-dimensional configurations we briefly comment on the M-theory setup of these solutions. 

The five-dimensional theory in question can also be obtained directly from eleven dimensions by using F-theory/M-theory duality. 
F-theory on $X\times S^1$ is dual to M-theory on $X$. The D3-brane wrapping ${\cal C} \times S^1$ with $n$ units of momentum is dual to an M2 brane wrapping a curve in the class $n[T^2] + [\cal C]$, where $\cal C$ is the curve in the base $B\subset X$, and $T^2$ is the elliptic fiber of $X$. In a type IIA setting, this is a bound state of $n$ D2-branes wrapping $T^2$ and one D2 wrapping ${\cal C}$.

Generically, M-theory on a Calabi-Yau manifold $X$ gives $n_V=h^{1,1}(X)-1$ vector multiplets and $n_H=h^{1,2}(X)+1$ hypermultiplets \cite{Cadavid:1995bk,Looyestijn:2010pb}. $h^{1,1}(X)$ vectors arise from expanding the eleven dimensional three-form in $H^2(X,\mathbb{R})$,
\begin{equation}\label{11d3-form}
C_{(3)}=\sum_{A=1}^{h^{1,1}(X)}A^A\wedge \omega_A+\dots\ ,
\end{equation}
where the ellipsis denote the terms leading to five-dimensional hypermultiplet scalars. One of these vector fields becomes the graviphoton while the others sit in the $n_V$ vector multiplets. The 5d real scalars in the vector multiplet correspond to the $h^{1,1}$ (uncomplexified) K\"ahler moduli of $X$. One combination, however, forms the volume modulus and sits in a hypermultiplet. These hypermultiplets play no role in our analysis, and are frozen to constant values. In the example of the elliptically fibered $X$ over base $B=\mathbb{P}^2$, we obtain 5d supergravity coupled to a single vector multiplet. The two one-forms from \eqref{11d3-form} correspond to $A$ and ${\tilde A}$ in \eqref{kk6-5}. They define the M2-brane charges, and their duals can support M5-brane flux. The radius field in \eqref{kk6-5} becomes the real scalar in the five-dimensional vector multiplet and measures the inverse area of the elliptic fiber (in 11d Planck units). For black hole solutions, the scalar is subject to the 5d attractor mechanism \cite{Ferrara:1996dd,Ferrara:1996um}, and we find at the horizon
\begin{equation}\label{radius}
\ex{2\varphi}|_{Hor.}=\sqrt{2}\, \frac{Q}{\tilde Q}\ .
\end{equation}
At infinity we have set $\ex{2\varphi}\to1$. (See Footnote~\ref{foot10}.) As an additional remark, we notice that we can truncate the 5d theory down to minimal supergravity. This can be done by choosing ${\cal F}=H$ such that the radius is constant everywhere, $e^{2\varphi}=1$. This can be achieved on BPS solutions by taking $H_1=H_5$ and $H_3=H_4$, such that, for single centers, $\mu=n$ and $q=p$ and hence ${\tilde Q}=\sqrt{2}Q$, consistent with \eqref{radius}. Bound state solutions we discuss below can therefore be truncated to minimal supergravity as well. Minimal supergravity can be embedded in F-theory by adding and freezing another 5d vector multiplet to the theory. If such a multiplet is not part of the spectrum, an F-theory embedding is not possible, as mentioned in Footnote \ref{foot1}. The M-theory picture does remain though.

The M-theory interpretation of the quantities \eqref{gauge inv quant} is given in Table~\ref{table relation M-theory}. Thus, the configurations of interest arise from configurations of KK-monopoles and antimonopoles, M2-branes and M5-brane charge fluxes.

\begin{table}[]
\begin{center}
\begin{tabular}{|c|c|c|c|} \hline
KK-monopole & M5-flux  &M2-charge & $\partial_{\psi}$\\  \hline  \rule{0pt}{2.5ex}  
$m_{a}$ &  $f_{a,a+1},\widetilde f_{a,a+1}$& $Q_{a},\widetilde Q_{a}$  & $J_{a}^{\psi}$ \\ \hline
\end{tabular}
\end{center}
\caption{Relation between residues in the functions $\mathbb H$ and charges of M-theory objects.}
\label{table relation M-theory}
\end{table}%

\subsection{Horizons and conserved charges}
\label{Physical properties app}

The dimensional reduction presented above allows one to obtain the full spectrum of five-dimensional BPS solutions in the time-like class \cite{Gauntlett:2002nw}: Reducing the solution \eqref{black string metric} along $u$ one obtains, depending on the value of $m$, the BMPV black hole \cite{Breckenridge:1996is} or the black hole with lens topology (black lens) recently discussed in \cite{Kunduri:2014kja}; the supersymmetric black ring \cite{Elvang:2004rt} can be obtained reducing a black tube solution. Finally there can be bound states of all these objects such as concentric black rings \cite{Gauntlett:2004wh,Gauntlett:2004qy}.

The nature of each object, characterized by the topology of the horizon $\mathcal H$, in a multicenter solution depends on the choice of charge vectors $\{\Gamma_{a}\}$ or, rather, the value of the gauge-invariant combinations \eqref{gauge inv quant}.  Schematically,\footnote{See Appendix~\ref{Near a center app} for a more detailed discussion.} the situation is  summarized as follows:
 \begin{enumerate}
 \item[{\it i.}]
 {\bf Black hole} (or black lens)
 $$ Q_a\tilde Q_a\neq0\,,\, m_a\neq0 \quad \implies  \quad  \mathcal H_{a}\sim S^3/\mathbb Z_{|m_a|} $$
 \item[{\it ii.}]
 {\bf Black ring}
 $$ Q_a\tilde Q_a\neq0\,, \,m_a=0 \quad \implies  \quad \mathcal H_{a}\sim S^1\times S^2 $$
 \item[{\it iii.}]
 {\bf Smooth center}
 $$ Q_a=\tilde Q_a=J_\psi{}_a=0\,, \, m_a\neq0 \qquad \implies \quad \text{no horizon}$$
 \end{enumerate}
The last case can be thought of a limit of the first case; the black hole horizon shrinks to zero size but it does so smoothly and the metric near such a GH center becomes $\mathbb{R}^{4}/\mathbb Z_{|m_a|}$. Here we focus our attention on configurations of bound states of black holes with finite $S^{3}$ horizons and, possibly, smooth centers in asymptotically flat $\Bbb R^{1,4}$. Rather surprisingly, these solutions have not been studied in the literature in detail. For  a study of black hole bound states in asymptotically AdS$_{3}\times S^{2}$ see \cite{deBoer:2008fk}. For the rest of the paper, unless otherwise specified, we will assume $m_a\neq0$. Although conical singularities are harmless in string theory, we take $|m_a|=1$ to have black holes with smooth $S^{3}$ horizons. Since asymptotic flatness requires $m_{\tot}=1$, we are necessarily led to consider ambipolar GH bases.

 In all the cases listed above, the entropy of the corresponding object is given by
 \equ{\label{black hole area}
S_a = \frac{\text{Area}(\mathcal H_{a})}{4\,G_5} =  \frac{2 \pi}{|m_a|} \sqrt{\frac{Q_a\tilde Q_a^2}2 - J_{\psi \,a}^2}\,,
}
where $G_5=\pi/4$ in our conventions. 

The charges $Q_a,\tilde Q_a$ and $J_{\psi a}$  in  \eqref{gauge inv quant}  can be given a geometric meaning, as the following integrals performed on the horizons $\mathcal H_a$:
\eqs{\nonumber
J_{\psi a}&=\frac1{4\pi^{2}} \int_{\mathcal H_a} \ast_{5} \,\rmd K^{(\psi)}\,, \\ \label{physical quantities centers}
 \tilde Q_a&=-\frac{\sqrt2}{8\pi^2} \int_{\mathcal H_a} \ast_{5} \, \widetilde F \,,\\
  Q_a&=-\frac1{8\pi^2} \int_{\mathcal H_a} \ast_{5} \,F  \,, \nonumber
}
where $K^{(\psi)}$ is the one-form associated to the Killing vector $\del_\psi$. Thus, $Q_a,\tilde Q_a$ and $J_{\psi a}$ are identified with the two electric charges and the angular momentum, respectively, of the black hole located at $\vec x=\vec x_a$. To interpret the quantities $f_{a,a+1}$ and $\widetilde{f}_{a,a+1}$ we recall the discussion below \eqref{H2 and eq chi}. In particular, the fact that a multicenter GH space contains $N-1$ independent 2-cycles, a basis for which is provided by the elements $\m C_{a,a+1}$ connecting the centers $x_a$ and $x_{a+1}$. Following \cite{Bena:2007kg} one can compute the fluxes of the magnetic part of the $U(1)$ field strengths $F$ and $\tilde F$, giving 
\equ{\label{fluxes general}
f_{a,a+1} =\frac1{4\pi} \int_{\m C_{a,a+1}} F \,,	\qquad \widetilde f_{a,a+1}=\frac{\sqrt2}{4\pi} \int_{\m C_{a,a+1}} \widetilde F  \,.
}
This also shows that all the combinations \eqref{gauge inv quant} can be expressed as integrals of gauge-invariant quantities.

In addition to the individual charges above, for asymptotically flat spacetimes one can define the total charges measured at infinity, given by the Komar integrals
\eqs{\label{physical quantities general}
J_{\psi}&=\frac1{4\pi^{2}} \int_{S^3_\infty} \ast_{5} \,\rmd K^{(\psi)} = 8( j\tot + \mu\tot p\tot + \frac12 n\tot q\tot + q\tot p\tot ^2)\,, \nn
 \widetilde Q&=-\frac{\sqrt2}{8\pi^2} \int_{S^3_\infty} \ast_{5} \,\rmd \widetilde A =4 \sqrt2( \mu\tot+p\tot q\tot) \,,\nn
  Q&=-\frac1{8\pi^2} \int_{S^3_\infty} \ast_{5} \,\rmd A =  4(n\tot+p\tot ^2) \,,}
 where $x\tot\equiv \sum_{a}x_{a}$ and we used $m\tot=1$ for asymptotically flat spacetimes. We note the the asymptotic charges are {\it not} simply given by the sum of the corresponding charges of each center. One can also define the total energy of the solution as the charge associated to the canonical time-like Killing vector $\del_t$:
\equ{\label{total mass}
M=-\frac3{32\pi^{2}} \int_{S^3_\infty} \ast_{5} \,\rmd K^{(t)} = \frac14\(Q+\sqrt2 \tilde Q\) \,,
}
where the second equality  is a consequence of the BPS condition. 

In the next section we will specialize the setting to multicenter configurations where all the centers lie on a single line, which we take to be the $z$-axis. As a consequence, the solution has an additional $U(1)$ isometry generated by rotations along the coordinate $\phi$. The corresponding angular momentum is given by
\equ{\label{extra angular momentum}
J_\phi=\frac1{32\pi^{2}} \int_{S^3_\infty} \ast_{5} \,\rmd K^{(\phi)} =\sum_a z_a\langle\Gamma_a,\Gamma_\infty\rangle\,,
}
where $z_a$ is the position of the $a$th center on the $z$-axis. For these axisymmetric solutions, the equations of motion \eqref{H2 and eq chi},  \eqref{eqs beta} and \eqref{eqs omega} can be easily solved for any number of centers. The expressions for the 1-forms $\bm \chi, \bm \beta,\bm \o$ in this case are given in Appendix~\ref{axysimmetric solution}.

\section{Bound states of two black holes in 5d}
\label{Bound states of black holes in 5d}

In this section, we study the simplest configuration of bound black holes with $S^{3}$ horizons in asymptotically flat $\Bbb R^{1,4}$ space. The configuration consists of two identical BMPV black holes and a smooth center, bound together by M5-charge flux through nontrivial topological cycles exterior to the horizons.  We shall show that, for fixed asymptotic charges, there is a small but finite region in parameter space where the twin black hole solutions exists and is regular, and even a region where it is entropically favored over the single-center black hole with the same asymptotic charges. 

\begin{figure}[]
\centering
\begin{tabular}{cc}
\includegraphics[page=1]{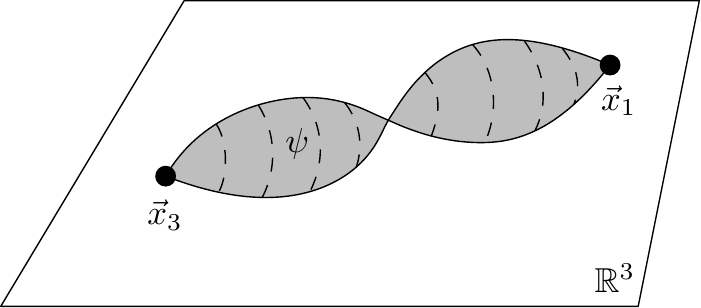}
\end{tabular}
\caption{A configuration of two black holes and a smooth center in the middle, bound by flux through the two 2-cycles of the GH base. }
\label{two BHs and a smooth center}
\end{figure}

Consider a three-center configuration with charge vectors $\Gamma_{a}$, $a=1,2,3$. Since we are interested in smooth horizons with $S^{3}$ topologies we take $m_{a}=\pm1$ and since asymptotic flatness requires $m_{\tot}=1$, the only choice is $m_{a}=(1,-1,1)$, up to trivial permutations. We take one of these centers, which we choose to be $\Gamma_{2}$, to be smooth, i.e.,   
\equ{
\widetilde{ Q}_{2}=  Q_{2}= J_{2}=0\,.
}
Thus, the charge vectors read
\equ{ \label{3 centers charges}
\Gamma_{1}=(\mu_1,1,0,0,n_1,j_{1})\,, \qquad \Gamma_{2}=(qp,-1,q,p,p^{2},\tfrac12 qp^{2})\,, \qquad \Gamma_{3}=(\mu_3,1,q_3,p_3,n_3,j_3)\,,
}
where we have set $p_1=q_1=0$ without loss of generality by the gauge transformation \eqref{shift residues}. If $\Gamma_3$ was chosen to be a smooth center, this would describe a single black hole and two smooth centers studied in \cite{Kunduri:2014iga}. Here, instead, we set $q_3=p_3=0$ and assume $Q_{1,3}, \tilde Q_{1,3}\neq0$, in which case the system corresponds to two BMPV black holes and a smooth center, as shown in Figure~\ref{two BHs and a smooth center}.

Explicitly, the harmonic functions are given by
\begin{align} \nonumber
H_{1}&=1+\frac{qp}{r}+\frac{\mu_1}{r_{1}}+\frac{\mu_3}{r_{3}} \,, & H_{2}&=-\frac{1}{r}+\frac{1}{r_{1}}+\frac{1}{r_3}\,, \\
H_{3}&=\frac{q}{r} \,, &  H_{4}&=\frac{p}{r}\,, \\ \nonumber
H_{5}&=1+\frac{p^{2}}{r}+\frac{n_1}{r_{1}}+\frac{n_3}{r_3} \,, &  H_{6}&=j_{\infty}+\frac{qp^{2}}{2r}+\frac{j_{1}}{r_{1}}+\frac{j_{3}}{r_3} \,,
\end{align}
where we have set $\vec x_{2}=0$ without loss of generality and $j_{\infty}=-p-\frac{q}2$ to ensure $\Bbb R^{1,4}$ asymptotics (see \ref{gamma inf r4 app}).

We emphasize that although the space is asymptotically $\mathbb R^{1,4}$ the GH base space is nontrivial so this is {\it not} a configuration of two BMPV black holes in $\mathbb R^{1,4}$. Instead, the geometry contains two nontrivial topological 2-cycles connecting each black hole to the smooth center.  As we discuss next, magnetic (or M5-charge) fluxes through these cycles keep the system bound.  

\subsection{Solving the bubble equations}

We now study the constraints on the relative locations of the GH centers imposed by the bubble equations. For simplicity, we focus on axisymmetric configurations, where the three GH centers lie on a straight line inside $\mathbb R^3$, which we take to be the $z$-axis and place the smooth center at the origin. We denote the locations of the two black holes by $\vec x_{1}=\{0,0,z_{1}\}$,  $\vec x_{3}=\{0,0,z_{3}\}$  and choose the orientation of the axis such that $z_1>0$.

The bubble equations \eqref{bubble eqs} read: 
\eqs{\nonumber
\frac{2\mu_1 p+n_1 q-p^{2}q-2j_{1}}{2\, z_1}+\frac{j_{1}-j_{3}}{|z_1-z_{3}|}=j_{\infty}\,, \\ \label{bubble eqs 3 centers}
\frac{2\mu_3 p+n_3 q-p^{2}q-2j_{3}}{2|z_{3}|}+\frac{j_{3}-j_{1}}{|z_1-z_{3}|}=j_{\infty}\,,
}
together with $j_{\infty}+p+\frac{q}2=0$, from \eqref{sumbubb}.

Let us assume for the moment that $j_{\infty}\neq 0$. We first note that in the case of identical twin black holes, i.e., $\Gamma_{1}=\Gamma_{3}\equiv(\mu,1,0,0,n,j)$, it follows from \eqref{bubble eqs 3 centers} that $z_3= \pm z_1$. Thus, to avoid Dirac-string singularities the two black holes must be either symmetrically located with respect to the origin, or  sit on top of each other. In the latter case, the configuration is in fact a two-center solution, consisting of a single black hole with horizon topology $S^{3}/\mathbb Z_{2}$ and a smooth center, studied in \cite{Kunduri:2014kja}. Here, instead, we consider the former case, for which 
\begin{equation}
\label{distances bound solution}
a\equiv z_1= -z_3=\frac{2j -nq +p^2q - 2 p\mu}{q+2p}\,.
\end{equation}
Thus, in this simple three-center configuration  the distances between the black holes and the smooth center are completely fixed.  Since $z_1>0$, for consistency we must require
\begin{equation}
\label{cond bubble eq 2bhs}
\frac{2j -nq +p^2q - 2 p\mu}{q+2p}>0~.
\end{equation}
This is the first restriction on the parameter space  $(\mu,q,p,n,j)$ characterizing the solution. Before analyzing additional constraints on parameter space, we comment briefly on scaling solutions. 

\paragraph{Comment on scaling solutions.}

Consider the special case  $j_{\infty} = 0$. We first note that if the black holes are identical, equations \eqref{bubble eqs 3 centers} in this case simply impose the relation
 \equ{ \label{relation charges unbound}
j -p(\mu+n-p^{2})=0\,,
}
but do not impose any constraints on $z_{1,3}$; as long as  \eqref{relation charges unbound} is satisfied, each black hole may sit anywhere along the $z$-axis without generating Dirac-string singularities.

If the black holes are not identical, however,  this is no longer the case. For instance, taking  $j_{1}\neq j_3$ it is easy to see that  consistency of  \eqref{bubble eqs 3 centers} requires $z_{3}>0$ and 
\equ{\label{ratio scaling solutions}
\frac{z_{1}}{z_{3}}=\frac{-j_{1}-np+p\mu+p^{3}}{j_{2}+np-p\mu-p^{3}}\,.
}
Thus, only the ratio $z_{1}/z_{3}$ is fixed, but not the overall scale of the system. For this class of solutions, one can define a scaling limit in which the asymptotically flat region of the metric decouples, resulting in an asymptotically  AdS$_{2}\times S^{3}$ space-time. Since the metric as one approaches any of the finite horizons is also  AdS$_{2}\times S^{3}$,  these solutions can be seen as interpolations between different AdS$_{2}\times S^{3}$ regions. These are usually referred to as {\it scaling} solutions  \cite{Bena:2006kb}.  Although interesting, we do not discuss this class  further, focusing instead on the case $j_{\infty}\neq 0$, describing truly bound solutions satisfying \eqref{distances bound solution}. 
 
\subsection{Spacetime regularity}
\label{Spacetime regularity}

In order for the solution to be physically acceptable, one must ensure the absence of closed timelike curves (CTCs) anywhere in spacetime \cite{Hawking:1991nk}. The absence of Dirac-string singularities discussed above is a necessary condition for the absence of CTCs close to the centers \cite{Bena:2005va}. However, as pointed out in \cite{Bena:2007kg}, this is not enough. For example, in order to avoid CTCs one must also impose
\begin{subequations}\label{expresions grr gtt gpsigpsi}
\eqs{\label{grr}
g_{rr}&=H_{2}f^{-1}>0\,,\\ \label{gpsipsi}
g_{\psi \psi}&=\frac{1}{f H_{2}}-f^{2}\omega_{\psi}^{2}>0\,.
}
\end{subequations}

A somewhat stronger requirement \cite{Berglund:2005vb,Hawking:1973uf}, which is sufficient to ensure the absence of CTCs globally, is to demand the metric to be stably causal and that $t$ provides a global time function. This is achieved if and only if
\equ{\label{gtt}
g^{tt}=-f^{-2}+f H_{2}\omega_{\psi}^{2}+f H_{2}^{-1}|\bm \omega_{i}|^{2}<0\,. 
}
Note that this condition does not necessarily imply both equations in \eqref{expresions grr gtt gpsigpsi}. If, however, $g_{rr}>0$ is satisfied then \eqref{gtt} does imply $g_{\psi\psi}>0$.

We will thus study the constraints imposed by \eqref{grr} and \eqref{gtt} on the parameters characterizing the configuration. We will show that the parameter space where these are satisfied is nonempty and, furthermore, that there is a region where it coexists with the single-center BMPV black hole with the same asymptotic charges and a region where the single-center solution would violate the CCB, but the multicenter is regular.  We carry out the analysis for the case of  twin black holes, located symmetrically from the smooth center at the distance \eqref{distances bound solution} as in Figure~\ref{two BHs and a smooth center}. We begin by analyzing \eqref{expresions grr gtt gpsigpsi} close to the centers. 

\paragraph{Close to the centers.} 

As discussed in Appendix~\ref{Near a center app}, close to the black hole horizons the regularity conditions $g_{rr}>0$ and $g^{tt}<0$ imply, respectively, 
\equ{ \label{conditions mu n j}
\mu, n>0\,, \qquad -\mu \sqrt{n}<j<\mu \sqrt{n}\,.
}
Note that since close to a center $\bm \o=0$  (see Appendix~\ref{Near a center app}) the condition \eqref{gtt} coincides with \eqref{gpsipsi}. These are the standard regularity conditions for a single-center BMPV black hole.  At the smooth centers the condition $g_{rr}>0$ reads  
 \equ{\label{grr close center} 
p^{2}>n + \frac12 a\,,  \qquad  pq> \mu +\frac12 a\,,
}
where we have used the fact that $z_{1}=-z_{3}\equiv a$  from \eqref{distances bound solution} to simplify the expressions. In addition, using the bubble equation \eqref{distances bound solution} one can see that close to the origin $-g^{tt}g_{rr}^{-2}=r^{2}+\mathcal O(r^{3})$. Thus, if \eqref{grr close center} holds, $g^{tt}<0$ is automatically satisfied close to the smooth center.

\paragraph{Away from the centers.}

We begin with the condition $g_{rr}>0$, which amounts to studying the positivity \eqref{grr} for all $r,\theta$. Remarkably, it is possible to prove that the positivity of this function close to the centers and at infinity is  sufficient to ensure this globally.  Precisely, under the assumption that the space is asymptotic to $\mathbb R^{1,4}$ and using the first inequalities in \eqref{conditions mu n j} and  \eqref{grr close center} one can prove that $H_{2}f^{-1}>0$ for any $r,\theta$. Thus,  
\equ{
\qquad \text{$g_{rr}>0$ at all centers} \qquad \Leftrightarrow  \qquad \text{$g_{rr}>0$ everywhere} \,.
}

We turn now to the study of possible CTCs, which amounts to studying \eqref{gtt} for all $r,\theta$. This is a rather nontrivial constraint and one should not expect the regularity conditions close to the centers  \eqref{conditions mu n j}, \eqref{grr close center} to be sufficient to ensure the absence of CTCs globally. In fact, there are known examples where this is not the case (see e.g. discussion in  \cite{Bena:2007kg}). However, within this class of solutions we have checked numerically in a vast number of instances of solutions to \eqref{conditions mu n j}, \eqref{grr close center}, that $g^{tt}<0$ is, in fact,  satisfied on the entire coordinate patch. Thus, the region in the five-dimensional parameter space $(\mu,q,p,n,j)$ where the configuration of twin back holes is globally regular (within our numerical analysis) is given by \eqref{cond bubble eq 2bhs}, \eqref{conditions mu n j} and \eqref{grr close center}. Before we analyze this parameter space in more detail, we discuss the physical properties of these solutions. 

\subsection{Global charges}

The mass, electric charges, and angular momenta of the twin black hole solution are given by the general expressions \eqref{physical quantities general}, \eqref{total mass}, \eqref{extra angular momentum} and read, in this particular case,
\begin{align}\label{totalcharges}
M &= 2\( n+p^2+2(\mu+pq)\),&& Q  = 8(n+p^2),&&	\tilde Q = 8\sqrt2 ( \mu + pq )\,,	\nn
J_\psi &= 8\(2j+n q+2 \mu p+3 p^2 q\)\,, && J_{\phi}=0\,.
\end{align}
We note that for this system $J_{\phi}=0$, which is a consequence of the bubble equations; computing the angular momentum \eqref{extra angular momentum} gives $J_\phi = \( p + \tfrac{q}{2} \) \(z_1 + z_3\)$, which vanishes  since  $z_1=-z_3$ as shown above.\footnote{We note that in the case of scaling solutions $2 p + q=0$ and hence $J_{\phi}=0$, even for nonidentical black holes.} Solutions with $J_{\phi}\neq 0$ can be obtained by considering nonidentical black holes.

In addition to the asymptotic charges, the solution is characterized by the fluxes \eqref{fluxes general} through the two independent 2-cycles $\mathcal C_{12}$ and $\mathcal C_{23}$ of the GH base:
\eqs{\label{fluxes 3-center}
f_{12}= f_{23}=q\,, \qquad \tilde f_{12}= \tilde f_{23}=\sqrt2\,p\,.
}
Recall that \eqref{grr close center} requires, in particular,  $pq>0$ and thus both fluxes must be nonzero for regularity.

\subsection{Parameter space and entropy}

Fixing the asymptotic charges does not completely specify the three-center solution. Indeed, one may solve say for  $n,\,\mu,\,q$ in terms of $Q,\,\tilde Q,\,J_\psi$, leaving the local quantities $j,p$ undetermined (although bounded by the regularity constraints). Thus, for a given set of asymptotic charges, there is a two-parameter family of twin black hole systems with the same asymptotics as the single-center solution. It is natural to compare the entropy of the two configurations. To first approximation, the entropy of the twin black hole system is simply given by the sum of the  Bekenstein-Hawking entropies of the two black holes, namely, 
\begin{align}\label{entropy}
S_{twin} &=32\pi\sqrt{\mu^2\,n - j^2} \nn
   &= \frac{32\pi}{Q}\sqrt{2\left[8 jp - \frac{ J_{\psi}}2 p +\frac{\tilde Q}{4\sqrt2}\(4p^{2}+\frac14 Q\)\right]^{2}\(\frac{Q}4-2p^{2}\)-Q^2\,j^{2}}\,,
\end{align}
where in the second line we have used \eqref{totalcharges} to write the expression in terms of asymptotic charges.  For comparison, the entropy of the single-center BMPV black hole with same asymptotic charges reads
\begin{equation}
S_\text{BMPV}=2\pi\sqrt{ \frac{\tilde Q^2 Q}2 - J_\psi^2}\,.
\end{equation}

At this point one can already see that for some choice of the asymptotic charges the twin black hole system can be entropically favored over the BMPV black hole. Indeed, consider a maximally-spinning BMPV black hole, i.e., 
\begin{equation}
|J_\psi|= \frac1{\sqrt2}\tilde Q \sqrt Q:\qquad \qquad S_\text{BMPV}=0 \qquad \text{and} \qquad S_{twin}>0\,,
\end{equation}
provided the requirements \eqref{conditions mu n j} and  \eqref{grr close center} are satisfied.

\medskip

We now discuss the region of existence of this solution in parameter space. Before discussing this we note  the entropy function has the following useful scaling properties
\begin{align}
&S(Q,\lambda\,\tilde Q, \lambda\, J_\psi,p, \lambda\, j) = \lambda\, S(Q,\,\tilde Q,  J_\psi,p,  j)\,,\\
&S(\Lambda^2Q,\,\tilde Q, \Lambda J_\psi,\Lambda p, \Lambda j) = \Lambda\, S(Q,\,\tilde Q,  J_\psi,p,  j)\,,
\end{align}
which can be used to write
\begin{equation}
S(Q,\tilde Q,  J_\psi,p,  j) = \tilde Q\sqrt{Q} \, S(1,1,\tfrac{J_\psi}{\tilde Q\sqrt{Q}}, \tfrac{p}{\sqrt{Q}}, \tfrac{j}{\tilde Q\sqrt{Q}})\,.
\end{equation}
Thus, it is convenient to define 
\begin{equation}
x = \tfrac{\sqrt2 J_\psi}{\tilde Q\sqrt{Q}}\,,\qquad \alpha= \tfrac{2p}{\sqrt{Q}}\,,\qquad \beta= \tfrac{8\sqrt2 j}{\tilde Q\sqrt{Q}}\,,
\end{equation}
in terms of which 
\begin{equation}
\frac {\sqrt2 S_{twin}}{\tilde Q\sqrt{Q}} \equiv \sigma(\alpha,\beta;x)= 4\pi \sqrt{\frac{1}{8} \left(1-2 \alpha ^2\right) (4 \alpha  (\alpha +2 \beta -x)+1)^2-\beta ^2} \label{entropy sigma}
\end{equation}
and
\begin{equation}
\frac {\sqrt2 S_\text{BMPV}}{\tilde Q\sqrt{Q}} =2\pi \sqrt{1-x^2}~.
\end{equation}
We note $x=1$ corresponds to the CCB for the single-center. \\

\begin{figure}[t]
\centering
\begin{tabular}{cc}
\includegraphics[width=2.1in]{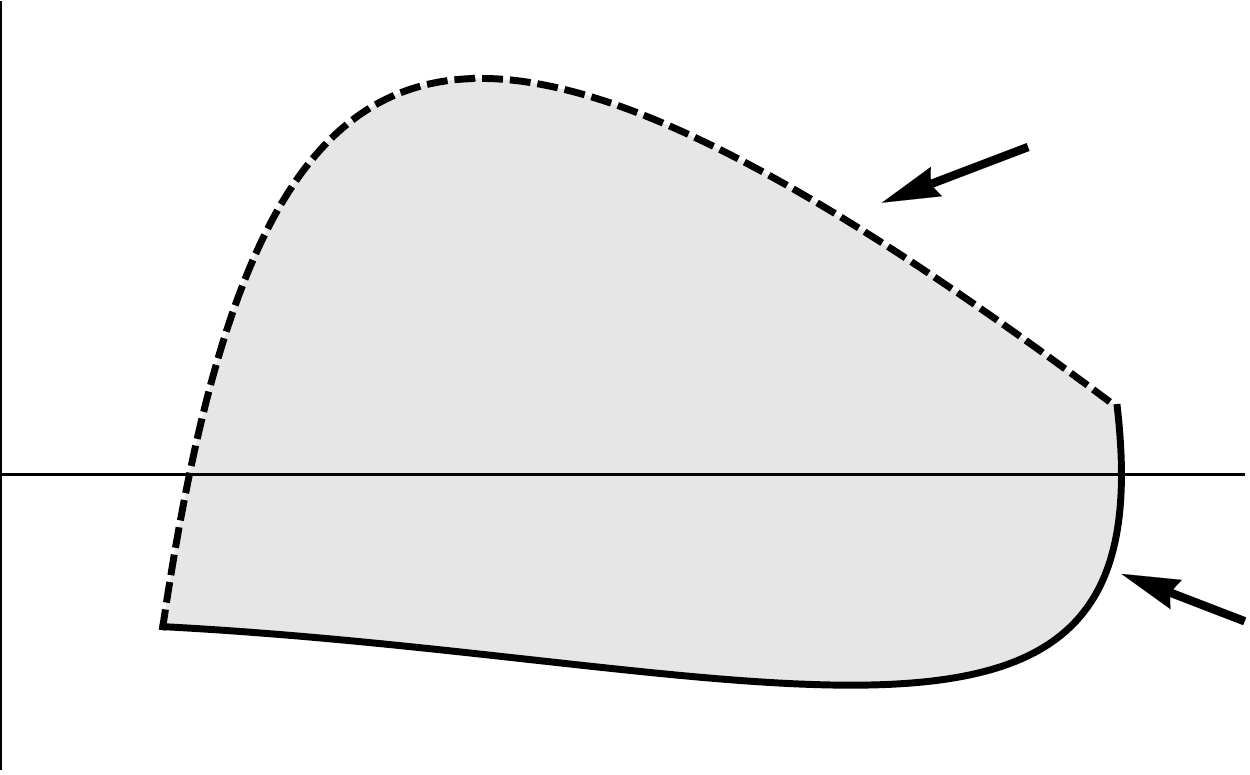} & \qquad \qquad \qquad \includegraphics[width=2.1in]{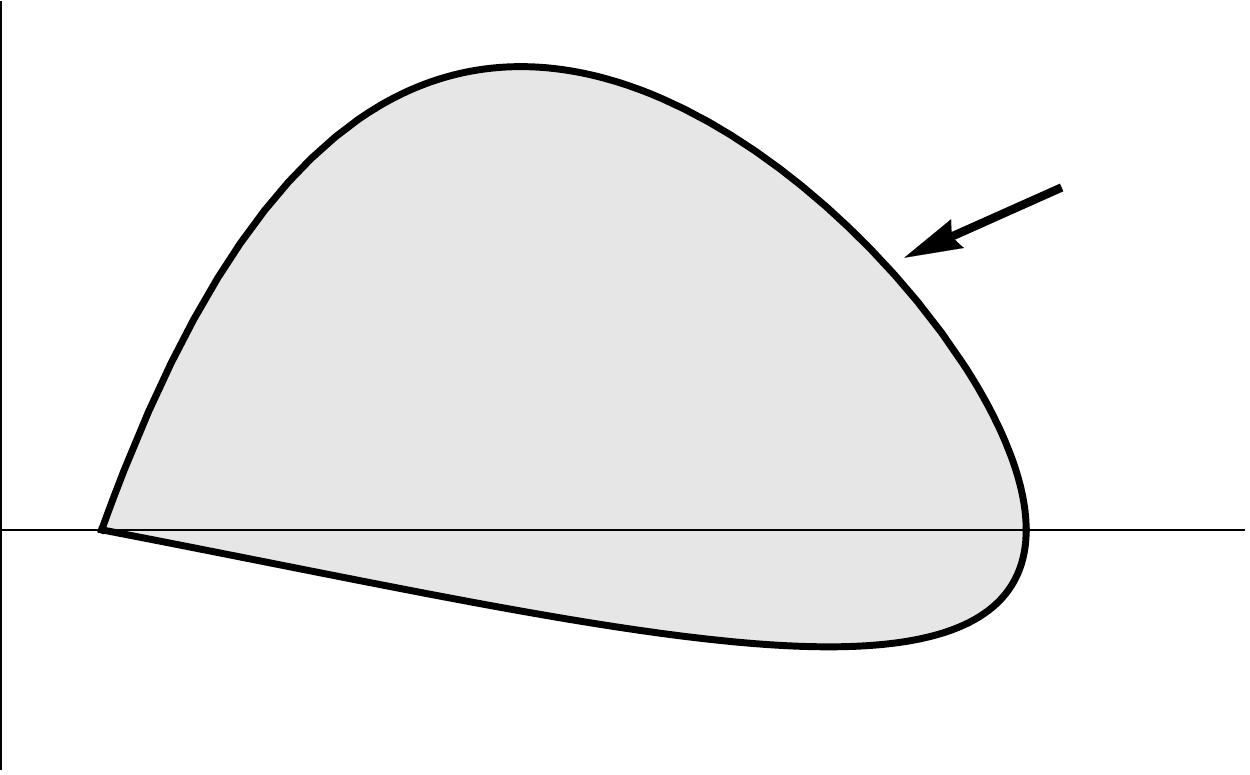}
\put(-2,30){ $\scriptstyle\alpha$}
\put(-236,36){ $\scriptstyle\alpha$}
\put(-395,98){ $\scriptstyle\beta$}
\put(-160,98){ $\scriptstyle\beta$}
\put(-235,15){ $\scriptstyle{g_{\psi\psi}^{(bh)}=0}$}
\put(-23,74){ $\scriptstyle{g_{\psi\psi}^{(bh)}=0}$}
\put(-262,77){ $\scriptstyle{a=0}$}
\end{tabular}
\caption{Region in $\alpha,\beta$ plane where the twin black hole solution is regular. For $x_{min}<x<1$ (left figure)  the single-center and twin black solutions coexist. For $1\leq x<x_{max}$ (right figure) the single-center solution violates the CCB but the twin black hole with the same asymptotic charges is regular.}
\label{entropyregion}
\end{figure}
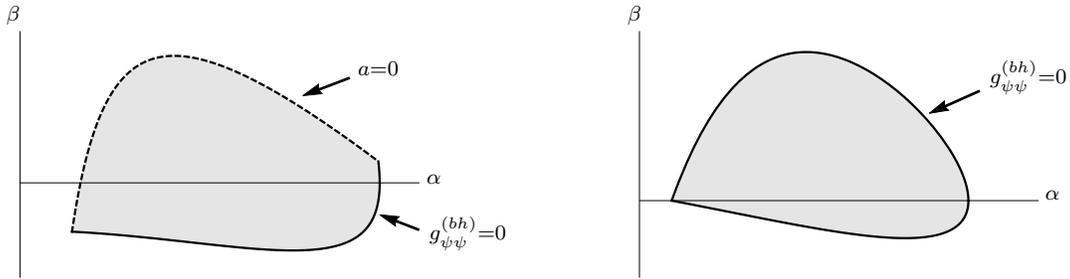

The five-dimensional parameter space of the twin black solution is spanned by the coordinates $(Q,\tilde Q,x,\alpha,\beta)$. To truly establish the existence and regularity of this solution, one must study whether there exists a region in this space where all the regularity conditions close to the centers \eqref{cond bubble eq 2bhs}, \eqref{conditions mu n j}, and \eqref{grr close center} are satisfied. As discussed in subsection \ref{Spacetime regularity} this is enough to also ensure global regularity in this class of solutions, at least within our numerical analysis of $g^{tt}$. It is easy to see that there exists a five-dimensional region in parameter space where all these conditions are satisfied and, in fact, it is infinite in extent. A more interesting question, however, is whether this is the case if the asymptotic charges $Q,\,\tilde Q,\,J_\psi$ are fixed, i.e., the two-dimensional slices of this space spanned by the parameters $\alpha, \beta$. The situation is slightly different in the cases $x<1$  and $x\geq1$, i.e., above and below the CCB of the single-center solution. The schematic situation is shown in Figure~\ref{entropyregion}. For $x_{min}<x<1$ with $x_{min} = \nicefrac{7\sqrt 2}{5\sqrt 5}$ the single-center and twin black solutions coexist. As shown in the left plot, the region where the twin black hole is regular is bounded on one side by the curve $a=0$, where the separation between the black holes goes to zero, and on the other side by the CCB for each individual black hole. 
For $1\leq x<x_{max}$ with $x_{max}=\nicefrac3{2\sqrt{2}}$ (right plot), the curve $g_{\psi\psi}=0$ closes on itself and entirely bounds the region of existence. For $x=x_{min}$ and $x=x_{max}$ the regions in Figure~\ref{entropyregion} shrink to a point. Although this figure suggests that the other regularity conditions, such as $g_{rr}>0$ or $g_{\psi\psi}>0$ close to the smooth center are unimportant, this is not the case; these are responsible for excluding other regions of parameter space which are not shown in the figure.

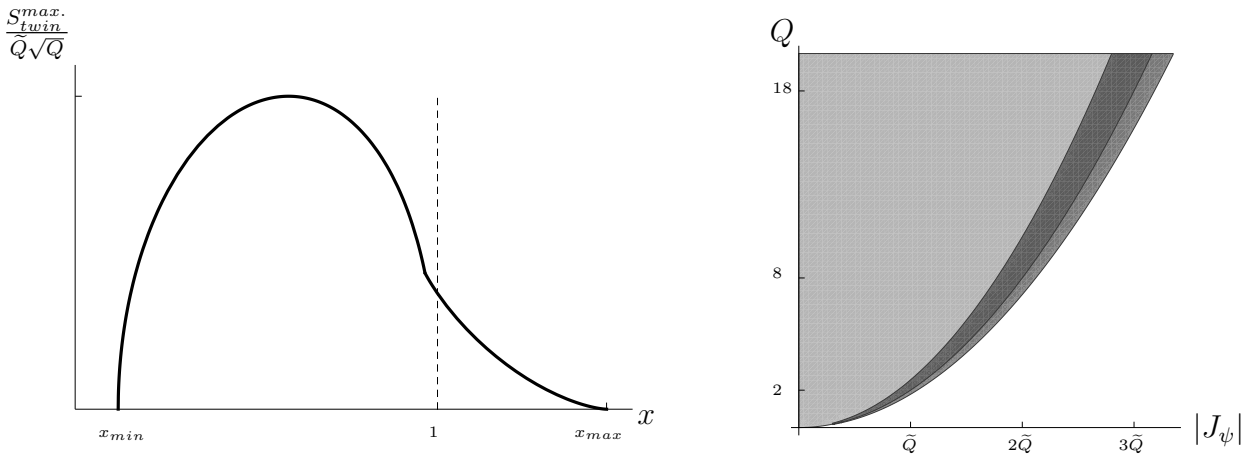
\begin{figure}[]
\centering
\begin{tabular}{cc}
\includegraphics[width=3in]{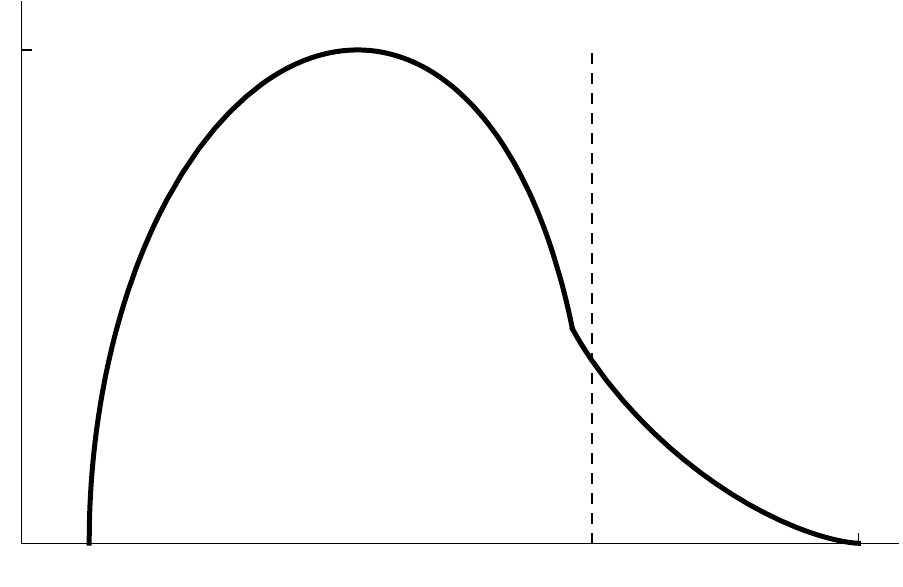} & \qquad \qquad \includegraphics[width=2.05in]{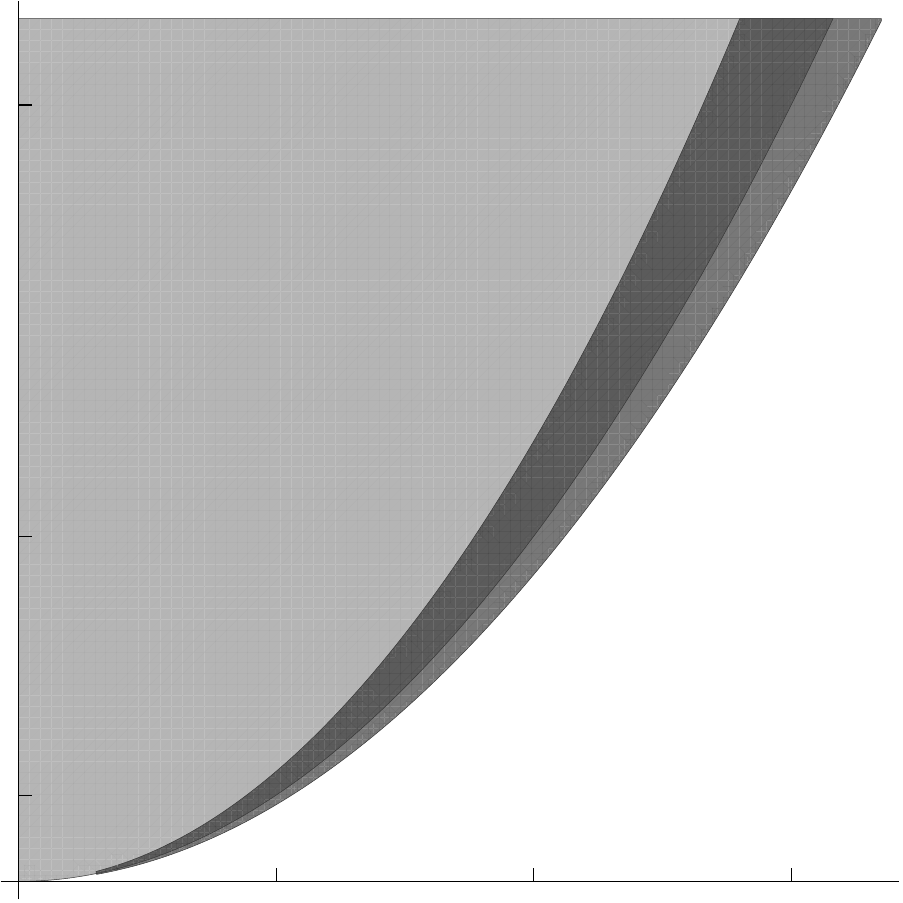}
\put(0,0){ $|J_{\psi}|$}
\put(-210,5){ $x$}
\put(-450,150){ $\tfrac{S_{twin}^{max.}}{\tilde Q \sqrt Q}$}
\put(-160,150){ $Q$}
\put(-410,0){\tiny $x_{min}$}
\put(-285,0){\tiny 1}
\put(-230,0){\tiny $x_{max}$}
\put(-110,-5){ \tiny $\tilde Q$}
\put(-70,-5){ \tiny $2\tilde Q$}
\put(-28,-5){ \tiny $3\tilde Q$}
\put(-155,16){\tiny 2}
\put(-155,60){\tiny 8}
\put(-155,130){\tiny 18}
\end{tabular}
\caption{Left plot: Maximum of the entropy as a function of $J_\psi$, for fixed values of $Q,\,\tilde Q$. Here   $x_{min} = \nicefrac{7\sqrt 2}{5\sqrt 5}$ and $x_{max}=\nicefrac3{2\sqrt{2}}$. The dashed vertical line represents the maximum value of the spin for the BMPV black hole. Right plot: Region where the twin black hole solution exists (dark gray) in the $(J_\psi,Q)$ plane, for a fixed value of $\tilde Q$. In the light gray region only the BMPV black hole exists. In the darkest region in the middle the two solutions coexist.}
\label{entropyplot}
\end{figure}

We now discuss the entropy of the twin black hole system. For any value of the parameters $\alpha, \beta$ belonging to the region shown in Figure~\ref{entropyregion} there exists a twin black hole solution with the same asymptotic charges as the single-center solution. Although the asymptotic charges are fixed, the entropy \eqref{entropy} of the twin black hole system depends on the values of $j, p$. To compare with the single-center entropy we may maximize $S_{twin}$ with respect to these quantities. Performing this maximization, making sure that the location of the maximum belongs to the regions in Figure~\ref{entropyregion}, one obtains the behavior displayed in the left plot in Figure~\ref{entropyplot}. For certain values of the the variable $x$, the maximum of the entropy is not an extremum, but in fact lies on the boundary of the allowed region; this is responsible for the kink in $S_{twin}^{max}$ in  Figure~\ref{entropyplot}.

\medskip

As shown in the right plot in Figure~\ref{entropyplot}, in $(Q,\tilde Q, J_\psi)$ space there is a narrow region where the twin black holes and the BMPV solution coexist. Namely, where the narrow dark strip overlaps with the interior of the gray region. The parabola represents the CCB for the single-center solution, where its entropy vanishes. In a small neighborhood to the left of the CCB both solutions exist and the twin black hole system has more entropy.\\

Of course, there is a vast number of multicenter solutions with the same asymptotic charges and some of these may be entropically favored over the twin back hole configuration discussed above. Indeed, one should expect that turning one of the twin black holes into a smooth center, distributing its charge into the other black hole, is entropically favored (see \cite{Kunduri:2014iga} for a study of this solution). In order to perform a full analysis of the ensemble defined by fixing the asymptotic charges, one should consider even more exotic configurations with arbitrary number of black rings, black holes, black lenses and smooth centers. This is beyond the aim of the present paper.

\section{Summary and Discussion}

We have shown the existence of a large class of regular, BPS black hole bound state solutions in asymptotically $\Bbb R^{1,4}$ spacetime. These were obtained by dimensional reduction of six-dimensional configurations of strings in minimal 6d (1,0) supergravity which admit an uplift to F-theory on an elliptically fibered Calabi-Yau threefold $X$ with base $\mathbb{P}^2$.  Other threefolds $X$ may also be considered, as long as the low-energy theory can be truncated to minimal 6d supergravity with frozen hypermultiplets. Presumably, these black hole bound states still exist in theories with additional supersymmetry, such as toroidal compactifications of M-theory or type IIB theory.

The particular multicenter configurations we have discussed consist of bound states of black holes with  $S^{3}$ horizons and smooth, horizonless, centers. We have shown that there is a finite region in parameter space where these configurations coexist with the single-center BMPV black hole with the same asymptotic charges, and a region where the single-center solution would violate the CCB, but the multicenter is regular.

Although a detailed analysis of the regularity constraints imposed on parameter space was carried out only in the three-center case (two black holes and a smooth center), one would expect similar results to hold in the case of an arbitrary number of centers. In particular, one may consider adding $N$ pairs of smooth centers with $m_{a}=-1$ and black holes with $m_{b}=1$ without changing the $\Bbb R^{1,4}$ asymptotics. 

Clearly, our results ask for a microscopic description. In F-theory, this might be possible using the recent results presented in \cite{Haghighat:2015ega}, though we should stress that the bound states described here are not the ones described in \cite{Haghighat:2015ega}. The latter are described by states with an entropy that is linear in the charges, and correspond to bound states of small black holes (with Taub-NUT asymptotics), while the bound states described in this paper can consist of large black holes. Moreover, we have additional fluxes that support the bound states which are absent in \cite{Haghighat:2015ega}. Presumably, the microscopics is easier to analyze in toroidal compactifications when the CFT has (4,4) supersymmetry rather than (0,4).

We have also discussed a set of $Sp(6,\Bbb R)$ transformations acting nontrivially on the solutions which, nonetheless, as a consequence of preserving the symplectic product, preserve the bubble equations. Subsets of these transformations have been considered in the literature before, including the generalized spectral flow transformations discussed and exploited in \cite{Bena:2008wt}.  It would be interesting to study the consequence of the full set of $Sp(6,\Bbb R)$ transformations \cite{CPV}.

Finally, we comment on the possibility of an ambipolar Taub-NUT base space, in which case the asymptotics in five dimensions would be $\Bbb R^{1,3}\times S^{1}$. This is interesting as the $S^{1}$ in the Taub-NUT geometry allows the further dimensional reduction down to four dimensions. The relation among various parameters that ensure the correct asymptotics of the six-dimensional solution are discussed in  Appendix~\ref{Asymptotics 6d app}. Although the setting is very similar to the asymptotically $\Bbb R^{1,4}$ case, to truly establish the existence and regularity of these solutions one must repeat the analysis performed above, which is left for future work. 

\bigskip
\bigskip

\noindent \textbf{Acknowledgements }

\bigskip

\noindent We  thank Nikolay Bobev, Nava Gaddam and Sameer Murthy for comments and feedback. The authors are supported by the Netherlands Organization for Scientific Research (NWO) under the VICI Grant 680-47-603. This work is part of the D-ITP consortium, a program of the NWO that is funded by the Dutch Ministry of Education, Culture and Science (OCW). 

\appendix

\section{6d solution and reduction to 5d}
\label{Six-dimensional metric}

In this Appendix, we provide some details of the six- and five-dimensional solutions discussed in the main text.
 
\subsection{The 1-forms $\chi, \beta, \o$}
\label{axysimmetric solution}

To fully specify the metric \eqref{BPSmetric5d} one must determine the 1-forms $\bm \chi, \bm \beta,\bm \o$ from equations \eqref{H2 and eq chi},  \eqref{eqs beta} and \eqref{eqs omega}, respectively. For simplicity,  we restrict ourselves to axisymmetric configurations, with all $N$ centers lying on a single line, which we take to be the $z$-axis, located at $\theta=(0,\pi)$. In this case it is easy to solve these equations, obtaining 
\equ{\label{multiGH}
\bm \chi = \sum_{a=1} ^{N}m_a \frac{r\cos(\theta)-z_a}{r_a}\,\rmd\phi  \,, \qquad \bm \beta =- \sum_{a=1} ^{N}q_a \frac{r\cos(\theta)-z_a}{r_a}\,\rmd\phi  \,,
}
and
\equ{\label{omegasol}
\bm \omega = \frac12\sum_{\substack{\{a,b\}=1\\a\neq b}}^{N}\langle \Gamma_a, \Gamma_b\rangle \frac{(z_a-z_b)^2 - (r_a-r_b)^2}{2(z_a-z_b)r_a r_b}\,\rmd\phi  + \sum_{a=1} ^{N}\langle \Gamma_\infty, \Gamma_a\rangle \frac{r\cos(\theta)-z_a}{r_a}\,\rmd\phi\,,
}
where $r_a \equiv|\vec x-\vec x_{a}|= \sqrt{r^2 + z_a^2 - 2\,r\,z_a\cos\theta}$. Of course, these 1-forms are defined up to gauge transformations; we have chosen a gauge in \eqref{omegasol} such that asymptotically $ \bm \omega \to  \langle \Gamma_\infty, \Gamma\tot\rangle \cos \theta \rmd \phi$, which vanishes  when \eqref{sumbubb} is imposed.

We note that evaluating \eqref{omegasol} on the (positive) $z$-axis gives 
\eqs{\label{omega axis}
\bm \omega|_{\theta=0}&=\sum_{a=1}^{N}\sign(r-z_{a})\(\langle \Gamma_{\infty},\Gamma_{a}\rangle-\sum_{b\neq a}\frac{\langle \Gamma_{a},\Gamma_{b}\rangle}{|z_{a}-z_{b}|}\)\rmd \phi\,.
}
Since the  function $\sign(r-z_{a})$ is discontinuous across the centers, the bubble equations \eqref{bubble eqs} ensure that $\bm \omega$ is regular (in fact,  vanishing) along the axis.\footnote{The expression on the negative $z$-axis is obtained by sending $r\to -r$ in this expression, and the same conclusion follows.}  

\subsection{Asymptotics}
\label{Asymptotics 6d app}

Requiring that asymptotically ($r\to \infty$) the functions $H, \mathcal F\to 1$ and the 1-forms $\o, \beta \to0$, the metric \eqref{6d metric general} asymptotes to $\mathbb R^1\times S^1_u\times X$, where $X$ is the asymptotics of the Gibbons-Hawking base.\footnote{\label{foot10}One can also relax the condition to $H^2\m F\to 1$ thus allowing for $H\to\mu_\infty$. This will rescale the radius of the asymptotic $S^1_u$. Here we fix $\mu_\infty=1$. } Setting $m_{\infty}=0$ and $m\tot=\sum_{a}m_{a}=1$ one has $X=\mathbb R^{4}$, while setting $m_{\infty}=1$ gives $X=\mathbb R^{3}\times S^{1}$.

In each case, this requirement leads to different relations among the parameters describing the solution. Setting $m_{\infty}=0$ and $m\tot=1$ and requiring that $H, \mathcal F\to 1$ and that $\o, \beta \to0$ one finds
\equ{
\mu_{\infty}=n_{\infty}=1\,, \qquad p_{\infty}=q_{\infty}=0\,.
}
Then the 1-forms are given asymptotically  by 
\equ{
 \beta = q\tot\rmd \psi+\mathcal O(r^{-1})\,, \qquad  \o = -\langle \Gamma_{\infty},\Gamma\tot\rangle \rmd \psi+\mathcal O(r^{-1})\,.
}
Imposing the bubble equation \eqref{sumbubb} implies $ \o$ identically vanishes at infinity in this gauge. The asymptotic form of $\beta$ can be absorbed by the simple coordinate redefinition $\rmd u\to \rmd u- q\tot \rmd \psi$. Note that, if the coordiante $u$ is periodic, then this is a change of coordinates on a torus and for it to be well-defined one must require
\equ{\label{q quantized}
\frac{\ell\,q\tot}{4\pi}  \in \mathbb Z~.
}
When the $u$-direction is compacified, this is a necessary condition for the space to be asymptotically $\mathbb R^{1,4}\times S^1_u$ with no conical defect. Imposing the constant term in $\o$ to vanish we obtain 
\equ{\label{gamma inf r4 app}
\Gamma_{\infty}=\(1,0,0,0,1, - p\tot-\frac{q\tot}2\)\,,
}
where $x\tot\equiv\sum_{a}x_{a}$.\\

 In the case of Taub-NUT, setting $m_{\infty}=1$ and imposing that $H, \mathcal F\to 1$ requires
\equ{
\mu_{\infty}+p_{\infty}q_{\infty}=1\,, \qquad n_{\infty}+p_{\infty}^{2}=1\,.
}
In this case, the 1-forms are given  asymptotically  by
\eqsn{
\beta&= q_{\infty} \, \rmd \psi+\( m\tot q_{\infty}- q\tot  \)\cos \theta \, \rmd \phi+\mathcal O(r^{-1})\,,\\
\o&= J_{\psi\infty} \, \rmd \psi +\(m\tot  J_{\psi\infty}  + \langle \Gamma_{\infty},\Gamma\tot\rangle\)\cos \theta \, \rmd \phi+\mathcal O(r^{-1})\,,
}
where  $ J_{\psi\infty}$ is as in \eqref{gauge inv quant}. Thus, if the bubble equation \eqref{sumbubb} is imposed one must set
\equ{
m\tot q_{\infty}=  q\tot \,, \qquad m\tot  J_{\psi\infty} =0\,.
}
Assuming $m\tot\neq 0$ the vector $\Gamma_{\infty}$ therefore reads 
\equ{\label{gamma inf TN app} 
\Gamma_{\infty}=\(1-\frac{p_{\infty} q\tot}{m\tot}, 1,\frac{q\tot}{m\tot},p_{\infty},1-p_{\infty}^{2},-p_{\infty}-\frac{q\tot(1-p_{\infty}^{2})}{2\,m\tot}\)\,.
}
The parameter $p_{\infty}$ is fixed in  terms of $\Gamma\tot$ from $\langle \Gamma_{\infty},\Gamma\tot\rangle=0$.

\subsection{Reduction to 5d}
\label{reduction}

The generic metric Ansatz for the reduction along a spacelike direction $u$  is
\eqs{\nonumber 
\rmd s^2_6 &= \ex{2\vf} \(\rmd u + A\)^2 + \ex{2\a\vf}\, \rmd s^2_5\,,\\ \label{kki}
\widehat G&=G+\frac12\rmd \widetilde A\wedge \(\rmd u + A\)\,,
}
where $\a$ is an arbitrary constant. Upon dimensional reduction along $u$ $\rmd s^2_5$ becomes the five-dimensional metric, $\vf$ becomes the dilaton field,  $A$ the graviphoton,  $G$ a 3-form field in five dimensions and $\tilde A$ another vector multiplet. The 6d Einstein-Hilbert Lagrangian, in terms of the 5d fields reads
\equ{
\sqrt{-g^{(6)}} \, R^{(6)} = \sqrt{g^{(5)}} \, \ex{(3\a + 1)\vf} \( R^{(5)} - 2 \,{\bf c}\, (\del\vf)^2 -\frac14 \ex{-2(\a-1)\vf} F^2\)\,,
}
where $F=\rmd A$ and ${\bf c} = (6\a^2 + 3\a + 1)$. If we demand the 5d Lagrangian to contain a canonical Einstein Hilbert term (Einstein frame) we must choose $\a = -1/3$. \\

We now apply this generic reduction to the solutions at hand. We assume the solution \eqref{6d metric general} is $u$-independent and reduce along this direction. The first step is to bring the metric into the form \eqref{kki}, namely:  
\equ{\label{6dBS}
\rmd s^{2}_{6}= H^{-1}\mathcal F\(\rmd u+\beta-\mathcal F^{-1}(\rmd v+\o)\)^{2}-H^{-1}\mathcal F^{-1}\(\rmd v+\o\)^{2}+H\, \rmd s^{2}_{4}\,.
}
Comparing this to \eqref{kki}, we identify the dilaton 
 \equ{\label{dilaton}
 \ex{2\vf} = H\inv \CF\,
 } 
and the five-dimensional metric in Einstein frame reads:
\equ{\label{metric 5d BH}
\rmd s^2_5 = -\frac{1}{(H^2 \CF)^{2/3}}\(\rmd t+w\)^2 + (H^2 \CF)^{1/3} \rmd s^2_4\,,
}
where we renamed $\rmd v =\rmd t$ and $t$ is a time-like direction in 5d. The graviphoton is given by 
\equ{\label{A 5d general}
A=\beta-\mathcal F^{-1}(\rmd t+\o)\,.
}
Now we turn to the reduction of the three-form field. In six dimensions, it is given by (see  \cite{Gutowski:2003rg})
\equ{\label{3-form in 6da}
\widehat G=\frac12 \ast_{4}\rmd H-\frac12 e^{+}\wedge (\rmd \omega)^{-}+\frac12 H^{-1}e^{-}\wedge \rmd \beta-\frac12 e^{+}\wedge e^{-}\wedge H^{-1}\rmd H\,,
}
where
\equ{
e^{+}=H^{-1}(\rmd u +\beta)\,, \qquad e^{-}=\rmd v+\omega -\frac{\m F}{2}(\rmd u+\beta)\,.
}
In order to bring this into the form form \eqref{kki} we perform some simple manipulations to write
\equ{\label{G for reduction 0}
\widehat G=-\frac12\(\rmd (H^{-1}(\rmd v+\omega))-\mathcal G^{+}\)\wedge \(\rmd u+A\)+\frac12 \ast_{4}\rmd H+\frac12 H^{-1}(\rmd v+\omega)\wedge\(\frac12 \rmd \beta-\m F^{-1}(\rmd \omega)^{-}\)
}
where we used $\rmd \omega^{-}=\tfrac12(\rmd \omega-\ast_{4}\rmd \omega)$ and defined
\equ{
\m G^{+}=H^{-1}\((\rmd \omega)^{+}-\frac12 \m{F} \, \rmd \beta \)\,.
}
For this class of solutions 
\equ{
\m G^{+}=\partial_{i}(H_{2}^{-1}H_{4})\rmd x^{i}\wedge \sigma-\frac12 \epsilon_{ijk} \partial_{k}(H_{2}^{-1}H_{4})H_{2}\rmd x^{i}\wedge \rmd x^{j}\,,
}
which can be written as
\equ{
\m G^{+}=\rmd (H_{2}^{-1}H_{4}\sigma)-\ast_{3} \rmd H_{4}=\rmd \(H_{2}^{-1}H_{4}\sigma+\bm \gamma\)
}
where in the second equality we introduced the 1-form $\bm \gamma$ satisfying
\equ{
\ast_{3}\rmd \bm \gamma= -\rmd  H_{4}\,.
}
Combining this with \eqref{G for reduction 0} we have
\eqs{\nonumber
\widehat G&=\frac12\, \rmd [-H^{-1}(\rmd v+\omega)+H_{2}^{-1}H_{4}\sigma+\bm \gamma]\wedge \(\rmd u+A\)\\
&+\frac12 \ast_{4}\rmd H+\frac12 H^{-1}(\rmd v+\omega)\wedge\(\frac12 \rmd \beta-\m F^{-1}(\rmd \omega)^{-}\)\,.
}
We may now compare this expression to  \eqref{kki} and identify the five-dimensional fields
\eqs{\label{AG 5d general}
\widetilde A&=-H^{-1}(\rmd t+\omega)+H_{2}^{-1}H_{4}\sigma+\bm \gamma\,,\\ \label{G 5d general}
G&=\frac12 \ast_{4}\rmd H+\frac12 H^{-1}(\rmd t+\omega)\wedge\(\frac12 \rmd \beta-\m F^{-1}(\rmd \omega)^{-}\)\,,
}
where $\rmd t=\rmd v$. Due to the self-duality of the 3-form $\widehat G$ in six dimensions, the 3-form $G$ in 5d is related to $\ast_{5}\rmd \widetilde A$ and is not an independent field. The field strengths of the two vector fields read:
\eqs{
F&= \rmd A = \mathcal F^{-2}\,\partial_{i}\mathcal F \, \rmd x^{i}\wedge (\rmd t+\o)+\rmd \beta -\mathcal F^{-1}\rmd \o\,,\\ \label{tilde F}
\widetilde F&=\rmd \widetilde A=-\rmd (H^{-1}(\rmd t+\omega))+\mathcal G^{+}\,.
}

\section{Five-dimensional metrics}
\label{Five-dimensional metric app}

Here we provide some details of the five-dimensional metrics studied in this paper. We discuss their asymptotics, their behavior close to the GH centers, and compute their physical properties (mass, electric charges, and magnetic fluxes) for solutions with an arbitrary number of centers. We also provide some details of the regularity constraints.

\subsection{Asymptotics}

Setting $\Gamma_{\infty}$ to \eqref{gamma inf r4 app}, and $m=1$, and reducing along the $u$-direction leads to a five-dimensional solution asymptotic to $\Bbb R\times \Bbb R^{4}$ while setting $\Gamma_{\infty}$  to \eqref{gamma inf TN app} (for $m\neq 0$)  leads to  $\Bbb R\times \Bbb R^{3}\times S^{1}$ asymptotics. 

Other interesting boundary conditions that one may consider in five dimensions are AdS$_{2}\times S^{3}$ or AdS$_{3}\times S^{2}$, obtained by setting  
\eqs{
\text{AdS}_{2}\times S^{3}&:\qquad \qquad \Gamma_{\infty}=0,\qquad m\tot=1\,,\\
\text{AdS}_{3}\times S^{2}&:\qquad \qquad \Gamma_{\infty}=0\,, \qquad m\tot=0\,.
} 

\subsection{Near a center}
\label{Near a center app}

Close to a center $\vec x_{a}$ the metric functions behave like:
\begin{align}
H&=H_{a}+ \frac{\widetilde{ Q}_{a}}{4\sqrt2m_{a}r_{a}}+\mathcal O(r_{a})\,,&  \mathcal F&=\mathcal F_{a}+\frac{ Q_{a}}{ 4m_{a}r_{a}}+\mathcal O(r_{a})\,, \\ \label{near center H,F}
  \omega _{\psi}&=\omega_{\psi}^{a}+ \frac{J_{a}^{\psi}}{8m_{a}^{2}r_{a}}+\mathcal O(r_{a})\,,&  \bm\omega_{i}&=\mathcal O(r_{a})\,,
\end{align}
where $\widetilde{ Q}_{a}, Q_{a},  J_{a}^{\psi}$ are given in \eqref{gauge inv quant} and $H_{a},\mathcal F_{a},\omega_{\psi}^{a}$ are constants which depend on the charges of all the centers, as well as their locations. The fact that $\bm \o$ vanishes to leading order is a consequence of the bubble equations (see e.g. \ref{omega axis}).  For a generic $\Gamma_{a}$ the $r_{a}^{-1}$  terms above are dominant. However, if the coefficients of these terms vanish, the leading behavior is controlled  by the constant parts; the metric behaves quite differently in these two cases. As we discuss now, in the former case the near-center metric coincides with (near-horizon) metric of a black hole with a finite-size horizon of topology $S^{3}/\Bbb Z_{|m_{a}|}$. In the latter case there is no horizon and the near-center metric is  $\Bbb R^{1,4}/\Bbb Z_{|m_{a}|}$.

\paragraph{Centers with $S^{3}$ horizons.}

Assuming $(\widetilde{  Q}_{a},   Q_{a}, J_{a}^{\psi})\neq0$ the $1/r_{a}$ parts in \eqref{near center H,F} are dominant and by a simple change of coordinates the metric near the center reads 
\equ{\label{metric near horizon}
\rmd s^{2}_{5}\simeq-\frac{r_{a}^{2}\rmd v^{2}}{\alpha_{a} m_{a}\lambda_{a}} \pm\frac{2 \alpha_{a} \rmd v \rmd r_{a}}{\sqrt{\alpha_{a}^{2}\lambda_{a}}}+\lambda_{a}\(\rmd \psi'+\chi_{\phi}^{(0)}\rmd \phi-\frac{r_{a} \sqrt{\alpha_{a}-m_{a}\lambda_{a}}}{\lambda_{a} \alpha_{a} \sqrt{m_{a}}}\rmd v \)^{2}+\alpha_{a} m_{a}\, (\rmd \theta^{2}+\sin^{2}\theta \rmd \phi^{2})\,,
}
where
\equ{\label{alpha and omega}
\alpha_{a}=\frac{1}{4m_a}\(\frac{\widetilde{ Q}_{a}^{2} Q_{a}}{2}\)^{1/3}\,, \qquad \lambda_{a} \equiv \frac{\alpha_{a}}{m_{a}}-\frac{ (J_{a}^{\psi})^{2}}{64m_{a}^{4}\alpha_{a}^{2}}\,.
}
The metric \eqref{metric near horizon} can  be shown to correspond to the near horizon of a BMPV black hole \cite{Breckenridge:1996is}, the two signs corresponding to a future or past horizon (see  \cite{Kunduri:2014iga} for a discussion). In particular the entropy associated to the center at $r_{a}=0$ is
\equ{\label{area horizon}
S_a=\frac{\mathcal A_{a}}{4 G_5}=\frac{1}{|m_{a}|}2\pi\sqrt{\frac{\widetilde{  Q}_{a}^{2} Q_{a}}2- J_{a}^{2}}\,.
}
In case of the single GH center with $m_{a}=1, q_{a}=p_{a}=0, j_{a}=j, \mu_{a}=\mu$ this becomes the entropy of the BMPV black hole $S=16\pi \sqrt{\mu^{2}n-j^{2}}$.

For the metric to be regular one must impose
\equ{
\widetilde{ Q}_{a}>0\,, \qquad   Q_{a}>0\,, \qquad \frac{\widetilde{Q}_{a}^{2} Q_{a}}2- (J_{a}^{\psi})^{2}>0\,.
}
In the case of a single center BMPV black hole the last condition above is the usual CCB.

\paragraph{Smooth horizonless centers.}

An interesting class of solutions are those for which
\equ{\label{conditions horizonless}
\widetilde{  Q}_{a}=Q_{a}= J_{a}^{\psi}=0\,.
}
In this case the functions \eqref{near center H,F} remain finite close to the center. For smooth centers the constant part $\omega_{\psi}^{a}$ is given by 
\equ{\label{omega0asmooth}
\omega_{\psi}^{a}= \frac{1}{m_{a}}\(\sum_{b\neq a}\frac{\langle \Gamma_{a}, \Gamma_{b} \rangle}{r_{ab}}-\langle \Gamma_{\infty},\Gamma_{a}\rangle\)\,,
}
which vanishes when the bubble equations are imposed. To avoid orbifold singularities in the base we set $m_{a}=\pm1$. Thus, close to a GH center satisfying  \eqref{conditions horizonless}, the metric is simply
\equ{\label{metric near smooth center}
\rmd s^2_{5}\simeq  -(H_{a}^{2}\mathcal F_{a})^{-2/3}\rmd t^2+(H_{a}^{2}\mathcal F_{a})^{1/3}\rmd s_{\pm \Bbb R^{4}}^2\,,
}
where $\rmd s_{\pm \Bbb R^{4}}^2=\pm \rmd s_{\Bbb R^{4}}^2$, where the two signs correspond to the sign of $m_{a}$. One can check that $\sign{(H_{a}^{2} \mathcal F_{a})}=\sign (m_{a})$, and thus the metric is smooth $\Bbb R^{1,4}$. In six dimensions, this uplifts to smooth $\Bbb R^{1,5}$. 

Finally, close to a center with $m_{a}=0$ the metric looks like the $r\to 0 $ limit of the single black string metric \eqref{black string metric}.

\bibliographystyle{utphys}
\bibliography{draft}

\providecommand{\href}[2]{#2}\begingroup\raggedright\begin{thebibliography}{10}

\bibitem{Breckenridge:1996is}
J.~C. Breckenridge, R.~C. Myers, A.~W. Peet, and C.~Vafa, ``{D-Branes and
  Spinning Black Holes},''
  \href{http://dx.doi.org/10.1016/S0370-2693(96)01460-8}{{\em Phys. Lett.} {\bf
  B391} (1997)  93--98},
\href{http://arxiv.org/abs/hep-th/9602065}{{\tt arXiv:hep-th/9602065
  [hep-th]}}.

\bibitem{Elvang:2004rt}
H.~Elvang, R.~Emparan, D.~Mateos, and H.~S. Reall, ``{A Supersymmetric Black
  Ring},'' \href{http://dx.doi.org/10.1103/PhysRevLett.93.211302}{{\em Phys.
  Rev. Lett.} {\bf 93} (2004)  211302},
\href{http://arxiv.org/abs/hep-th/0407065}{{\tt arXiv:hep-th/0407065
  [hep-th]}}.

\bibitem{Kunduri:2014kja}
H.~K. Kunduri and J.~Lucietti, ``{Supersymmetric Black Holes with Lens-Space
  Topology},'' \href{http://dx.doi.org/10.1103/PhysRevLett.113.211101}{{\em
  Phys. Rev. Lett.} {\bf 113} (2014) no.~21, 211101},
\href{http://arxiv.org/abs/1408.6083}{{\tt arXiv:1408.6083 [hep-th]}}.

\bibitem{Gibbons:2013tqa}
G.~W. Gibbons and N.~P. Warner, ``{Global structure of five-dimensional
  fuzzballs},'' \href{http://dx.doi.org/10.1088/0264-9381/31/2/025016}{{\em
  Class. Quant. Grav.} {\bf 31} (2014)  025016},
\href{http://arxiv.org/abs/1305.0957}{{\tt arXiv:1305.0957 [hep-th]}}.

\bibitem{Gauntlett:2004wh}
J.~P. Gauntlett and J.~B. Gutowski, ``{Concentric Black Rings},''
  \href{http://dx.doi.org/10.1103/PhysRevD.71.025013}{{\em Phys. Rev.} {\bf
  D71} (2005)  025013},
\href{http://arxiv.org/abs/hep-th/0408010}{{\tt arXiv:hep-th/0408010
  [hep-th]}}.

\bibitem{Gauntlett:2004qy}
J.~P. Gauntlett and J.~B. Gutowski, ``{General Concentric Black Rings},''
  \href{http://dx.doi.org/10.1103/PhysRevD.71.045002}{{\em Phys. Rev.} {\bf
  D71} (2005)  045002},
\href{http://arxiv.org/abs/hep-th/0408122}{{\tt arXiv:hep-th/0408122
  [hep-th]}}.

\bibitem{Bena:2007kg}
I.~Bena and N.~P. Warner, ``{Black Holes, Black Rings and Their Microstates},''
  \href{http://dx.doi.org/10.1007/978-3-540-79523-0_1}{{\em Lect. Notes Phys.}
  {\bf 755} (2008)  1--92},
\href{http://arxiv.org/abs/hep-th/0701216}{{\tt arXiv:hep-th/0701216
  [hep-th]}}.

\bibitem{Vafa:1997gr}
C.~Vafa, ``{Black Holes and Calabi-Yau Threefolds},'' {\em Adv. Theor. Math.
  Phys.} {\bf 2} (1998)  207--218,
\href{http://arxiv.org/abs/hep-th/9711067}{{\tt arXiv:hep-th/9711067
  [hep-th]}}.

\bibitem{Haghighat:2015ega}
B.~Haghighat, S.~Murthy, C.~Vafa, and S.~Vandoren, ``{F-Theory, Spinning Black
  Holes and Multi-String Branches},''
  \href{http://dx.doi.org/10.1007/JHEP01(2016)009}{{\em JHEP} {\bf 01} (2016)
  009},
\href{http://arxiv.org/abs/1509.00455}{{\tt arXiv:1509.00455 [hep-th]}}.

\bibitem{Gauntlett:1998fz}
J.~P. Gauntlett, R.~C. Myers, and P.~K. Townsend, ``{Black holes of D = 5
  supergravity},'' \href{http://dx.doi.org/10.1088/0264-9381/16/1/001}{{\em
  Class. Quant. Grav.} {\bf 16} (1999)  1--21},
\href{http://arxiv.org/abs/hep-th/9810204}{{\tt arXiv:hep-th/9810204
  [hep-th]}}.

\bibitem{Gauntlett:2002nw}
J.~P. Gauntlett, J.~B. Gutowski, C.~M. Hull, S.~Pakis, and H.~S. Reall, ``{All
  supersymmetric solutions of minimal supergravity in five- dimensions},''
  \href{http://dx.doi.org/10.1088/0264-9381/20/21/005}{{\em Class. Quant.
  Grav.} {\bf 20} (2003)  4587--4634},
\href{http://arxiv.org/abs/hep-th/0209114}{{\tt arXiv:hep-th/0209114
  [hep-th]}}.

\bibitem{Bena:2011zw}
I.~Bena, B.~D. Chowdhury, J.~de~Boer, S.~El-Showk, and M.~Shigemori,
  ``{Moulting Black Holes},''
  \href{http://dx.doi.org/10.1007/JHEP03(2012)094}{{\em JHEP} {\bf 03} (2012)
  094},
\href{http://arxiv.org/abs/1108.0411}{{\tt arXiv:1108.0411 [hep-th]}}.

\bibitem{Kunduri:2014iga}
H.~K. Kunduri and J.~Lucietti, ``{Black Hole Non-Uniqueness via Spacetime
  Topology in Five Dimensions},''
  \href{http://dx.doi.org/10.1007/JHEP10(2014)082}{{\em JHEP} {\bf 10} (2014)
  82},
\href{http://arxiv.org/abs/1407.8002}{{\tt arXiv:1407.8002 [hep-th]}}.

\bibitem{Vafa:1996xn}
C.~Vafa, ``{Evidence for F Theory},''
  \href{http://dx.doi.org/10.1016/0550-3213(96)00172-1}{{\em Nucl. Phys.} {\bf
  B469} (1996)  403--418},
\href{http://arxiv.org/abs/hep-th/9602022}{{\tt arXiv:hep-th/9602022
  [hep-th]}}.

\bibitem{Morrison:1996na}
D.~R. Morrison and C.~Vafa, ``{Compactifications of F Theory on Calabi-Yau
  Threefolds. 1},'' \href{http://dx.doi.org/10.1016/0550-3213(96)00242-8}{{\em
  Nucl. Phys.} {\bf B473} (1996)  74--92},
\href{http://arxiv.org/abs/hep-th/9602114}{{\tt arXiv:hep-th/9602114
  [hep-th]}}.

\bibitem{Morrison:1996pp}
D.~R. Morrison and C.~Vafa, ``{Compactifications of F Theory on Calabi-Yau
  Threefolds. 2.},'' \href{http://dx.doi.org/10.1016/0550-3213(96)00369-0}{{\em
  Nucl. Phys.} {\bf B476} (1996)  437--469},
\href{http://arxiv.org/abs/hep-th/9603161}{{\tt arXiv:hep-th/9603161
  [hep-th]}}.

\bibitem{Gutowski:2003rg}
J.~B. Gutowski, D.~Martelli, and H.~S. Reall, ``{All Supersymmetric Solutions
  of Minimal Supergravity in Six- Dimensions},''
  \href{http://dx.doi.org/10.1088/0264-9381/20/23/008}{{\em Class. Quant.
  Grav.} {\bf 20} (2003)  5049--5078},
\href{http://arxiv.org/abs/hep-th/0306235}{{\tt arXiv:hep-th/0306235
  [hep-th]}}.

\bibitem{Bena:2006qm}
I.~Bena, D.-E. Diaconescu, and B.~Florea, ``{Black String Entropy and
  Fourier-Mukai Transform},''
  \href{http://dx.doi.org/10.1088/1126-6708/2007/04/045}{{\em JHEP} {\bf 04}
  (2007)  045},
\href{http://arxiv.org/abs/hep-th/0610068}{{\tt arXiv:hep-th/0610068
  [hep-th]}}.

\bibitem{Bena:2011dd}
I.~Bena, S.~Giusto, M.~Shigemori, and N.~P. Warner, ``{Supersymmetric Solutions
  in Six Dimensions: A Linear Structure},''
  \href{http://dx.doi.org/10.1007/JHEP03(2012)084}{{\em JHEP} {\bf 03} (2012)
  084},
\href{http://arxiv.org/abs/1110.2781}{{\tt arXiv:1110.2781 [hep-th]}}.

\bibitem{Gibbons:1979zt}
G.~W. Gibbons and S.~W. Hawking, ``{Gravitational Multi - Instantons},''
\href{http://dx.doi.org/10.1016/0370-2693(78)90478-1}{{\em Phys. Lett.} {\bf
  B78} (1978)  430}.

\bibitem{Bena:2005va}
I.~Bena and N.~P. Warner, ``{Bubbling Supertubes and Foaming Black Holes},''
  \href{http://dx.doi.org/10.1103/PhysRevD.74.066001}{{\em Phys. Rev.} {\bf
  D74} (2006)  066001},
\href{http://arxiv.org/abs/hep-th/0505166}{{\tt arXiv:hep-th/0505166
  [hep-th]}}.

\bibitem{Berglund:2005vb}
P.~Berglund, E.~G. Gimon, and T.~S. Levi, ``{Supergravity Microstates for BPS
  Black Holes and Black Rings},''
  \href{http://dx.doi.org/10.1088/1126-6708/2006/06/007}{{\em JHEP} {\bf 06}
  (2006)  007},
\href{http://arxiv.org/abs/hep-th/0505167}{{\tt arXiv:hep-th/0505167
  [hep-th]}}.

\bibitem{Saxena:2005uk}
A.~Saxena, G.~Potvin, S.~Giusto, and A.~W. Peet, ``{Smooth Geometries with Four
  Charges in Four Dimensions},''
  \href{http://dx.doi.org/10.1088/1126-6708/2006/04/010}{{\em JHEP} {\bf 04}
  (2006)  010},
\href{http://arxiv.org/abs/hep-th/0509214}{{\tt arXiv:hep-th/0509214
  [hep-th]}}.

\bibitem{Bobev:2012af}
N.~Bobev, B.~E. Niehoff, and N.~P. Warner, ``{New Supersymmetric Bubbles on
  AdS$_3xS^3$},'' \href{http://dx.doi.org/10.1007/JHEP10(2012)013}{{\em JHEP}
  {\bf 10} (2012)  013},
\href{http://arxiv.org/abs/1204.1972}{{\tt arXiv:1204.1972 [hep-th]}}.

\bibitem{Denef:2000nb}
F.~Denef, ``{Supergravity Flows and D-Brane Stability},''
  \href{http://dx.doi.org/10.1088/1126-6708/2000/08/050}{{\em JHEP} {\bf 08}
  (2000)  050},
\href{http://arxiv.org/abs/hep-th/0005049}{{\tt arXiv:hep-th/0005049
  [hep-th]}}.

\bibitem{Bena:2008wt}
I.~Bena, N.~Bobev, and N.~P. Warner, ``{Spectral Flow, and the Spectrum of
  Multi-Center Solutions},''
  \href{http://dx.doi.org/10.1103/PhysRevD.77.125025}{{\em Phys. Rev.} {\bf
  D77} (2008)  125025},
\href{http://arxiv.org/abs/0803.1203}{{\tt arXiv:0803.1203 [hep-th]}}.

\bibitem{Bena:2005ni}
I.~Bena, P.~Kraus, and N.~P. Warner, ``{Black Rings in Taub-Nut},''
  \href{http://dx.doi.org/10.1103/PhysRevD.72.084019}{{\em Phys. Rev.} {\bf
  D72} (2005)  084019},
\href{http://arxiv.org/abs/hep-th/0504142}{{\tt arXiv:hep-th/0504142
  [hep-th]}}.

\bibitem{CPV}
M.~Crichigno, F.~Porri, and S.~Vandoren {\em To appear}  .

\bibitem{Cadavid:1995bk}
A.~C. Cadavid, A.~Ceresole, R.~D'Auria, and S.~Ferrara, ``{Eleven-Dimensional
  Supergravity Compactified on Calabi-Yau Threefolds},''
  \href{http://dx.doi.org/10.1016/0370-2693(95)00891-N}{{\em Phys. Lett.} {\bf
  B357} (1995)  76--80},
\href{http://arxiv.org/abs/hep-th/9506144}{{\tt arXiv:hep-th/9506144
  [hep-th]}}.

\bibitem{Looyestijn:2010pb}
H.~Looyestijn, E.~Plauschinn, and S.~Vandoren, ``{New Potentials from
  Scherk-Schwarz Reductions},''
  \href{http://dx.doi.org/10.1007/JHEP12(2010)016}{{\em JHEP} {\bf 12} (2010)
  016},
\href{http://arxiv.org/abs/1008.4286}{{\tt arXiv:1008.4286 [hep-th]}}.

\bibitem{Ferrara:1996dd}
S.~Ferrara and R.~Kallosh, ``{Supersymmetry and attractors},''
  \href{http://dx.doi.org/10.1103/PhysRevD.54.1514}{{\em Phys. Rev.} {\bf D54}
  (1996)  1514--1524},
\href{http://arxiv.org/abs/hep-th/9602136}{{\tt arXiv:hep-th/9602136
  [hep-th]}}.

\bibitem{Ferrara:1996um}
S.~Ferrara and R.~Kallosh, ``{Universality of supersymmetric attractors},''
  \href{http://dx.doi.org/10.1103/PhysRevD.54.1525}{{\em Phys. Rev.} {\bf D54}
  (1996)  1525--1534},
\href{http://arxiv.org/abs/hep-th/9603090}{{\tt arXiv:hep-th/9603090
  [hep-th]}}.

\bibitem{deBoer:2008fk}
J.~de~Boer, F.~Denef, S.~El-Showk, I.~Messamah, and D.~Van~den Bleeken,
  ``{Black hole bound states in AdS(3) x S**2},''
  \href{http://dx.doi.org/10.1088/1126-6708/2008/11/050}{{\em JHEP} {\bf 11}
  (2008)  050},
\href{http://arxiv.org/abs/0802.2257}{{\tt arXiv:0802.2257 [hep-th]}}.

\bibitem{Bena:2006kb}
I.~Bena, C.-W. Wang, and N.~P. Warner, ``{Mergers and typical black hole
  microstates},'' \href{http://dx.doi.org/10.1088/1126-6708/2006/11/042}{{\em
  JHEP} {\bf 11} (2006)  042},
\href{http://arxiv.org/abs/hep-th/0608217}{{\tt arXiv:hep-th/0608217
  [hep-th]}}.

\bibitem{Hawking:1991nk}
S.~W. Hawking, ``{The Chronology Protection Conjecture},''
\href{http://dx.doi.org/10.1103/PhysRevD.46.603}{{\em Phys. Rev.} {\bf D46}
  (1992)  603--611}.

\bibitem{Hawking:1973uf}
S.~W. Hawking and G.~F.~R. Ellis,
  \href{http://dx.doi.org/10.1017/CBO9780511524646}{{\em {The Large Scale
  Structure of Space-Time}}}.
\newblock Cambridge Monographs on Mathematical Physics. Cambridge University
  Press,
2011.
\newblock

\end{thebibliography}\endgroup

\end{document}